\documentclass[journal]{IEEEtran}

\ifCLASSINFOpdf
\else
\fi
\usepackage[T1]{fontenc}
\usepackage[latin9]{inputenc}
\usepackage{float}
\usepackage{calc}
\usepackage{amsmath}
\usepackage{graphicx}
\usepackage{mathtools}
\usepackage[export]{adjustbox}
\usepackage{listings}
\usepackage{amssymb}
\usepackage{flexisym}
\usepackage{xcolor}
\usepackage{wrapfig}
\usepackage{textcase}
\usepackage{soul}
\usepackage{comment}
\usepackage{multirow}
\usepackage{lmodern}

\DeclareMathOperator*{\argmin}{arg\,min}

\usepackage[ruled]{algorithm2e}
\usepackage{tabularx,booktabs}
\usepackage[numbers,sort&compress]{natbib}
\usepackage{tikz}
\usetikzlibrary{positioning}
\usepackage{nameref}
\usepackage{dblfloatfix}

\SetKwInOut{KwInput}{Input}                
\SetKwInOut{KwOutput}{Output} 

\begin{document}

\title{Fast Hyperspectral Neutron Tomography}

\author{Mohammad Samin Nur Chowdhury$^{1}$, Diyu Yang$^{2}$, Shimin Tang$^{3}$, Singanallur V. Venkatakrishnan$^{4}$, Hassina Z. Bilheux$^{3}$, Gregery T. Buzzard$^{5}$, and Charles A. Bouman$^{1}$\\
$^{1}$School of Electrical and Computer Engineering, Purdue University, IN 47907, USA\\
$^{2}$Apple Inc., CA 95014, USA\\
$^{3}$Neutron Scattering Division, Oak Ridge National Laboratory, TN 37830, USA\\ 
$^{4}$Electrical and Engineering Infrastructure Division, Oak Ridge National Laboratory, TN 37831, USA\\ 
$^{5}$Department of Mathematics, Purdue University, IN 47907, USA
\thanks{This manuscript has been authored by UT-Battelle, LLC, under contract DE-AC05-00OR22725 with the US Department of Energy (DOE). The US government retains and the publisher, by accepting the article for publication, acknowledges that the US government retains a nonexclusive, paid-up, irrevocable, worldwide license to publish or reproduce the published form of this manuscript, or allow others to do so, for US government purposes. DOE will provide public access to these results of federally sponsored research in accordance with the DOE Public Access Plan (http://energy.gov/downloads/doe-public-access-plan).}\\}

\maketitle

\begin{abstract}

Hyperspectral neutron computed tomography is a tomographic imaging technique in which thousands of wavelength-specific neutron radiographs are measured for each tomographic view.
In conventional hyperspectral reconstruction, data from each neutron wavelength bin are reconstructed separately, which is extremely time-consuming.
These reconstructions often suffer from poor quality due to low signal-to-noise ratios.
Consequently, material decomposition based on these reconstructions tends to produce inaccurate estimates of the material spectra and erroneous volumetric material separation.
In this paper, we present two novel algorithms for processing hyperspectral neutron data: 
fast hyperspectral reconstruction and fast material decomposition. 
Both algorithms rely on a subspace decomposition procedure that transforms hyperspectral views into low-dimensional projection views within an intermediate subspace, where tomographic reconstruction is performed.
The use of subspace decomposition dramatically reduces reconstruction time while reducing both noise and reconstruction artifacts.
We apply our algorithms to both simulated and measured neutron data and demonstrate that they reduce computation and improve the quality of the results relative to conventional methods.

\end{abstract}


\IEEEpeerreviewmaketitle

\section{Introduction}
\label{sec:introduction}

\IEEEPARstart{N}{eutron} computed tomography (nCT) can reveal an object's internal structure from exposures to a neutron source at multiple orientations.
Unlike X-rays, neutrons interact directly with atomic nuclei rather than electron clouds, making nCT particularly useful for studying materials that are challenging to analyze with X-ray methods \cite{vontobel2005nctvsxct, solorzano2014nctvsxct}. 
nCT is also valuable for detecting contaminants in materials, such as nuclear fuel elements or composite structures \cite{kim1999composite}, and for monitoring aging in polymers, metals, and other materials, revealing how environmental factors affect material longevity \cite{chin2007aging}.
Additionally, nCT is used in biological imaging \cite{bilheux2014biology}, archaeological research \cite{kardjilov2006archeology}, and manufacturing quality control \cite{garbe2017industrial}.

Hyperspectral neutron computed tomography (HSnCT) is a more advanced technique \cite{tran2021hyperspectral, venkat2019hyperspectral, venkat2021hyperspectral}, in which a pulsed neutron source illuminates a sample and a time-of-flight detector measures the projection images across a broad range of wavelengths - potentially of the order of few thousands.
Using HSnCT, it is possible to analyze material characteristics like crystallographic phases \cite{ametova2021thermal, woracek2014thermal} and isotopic compositions \cite{balke2024epithermal, balke2021epithermal, losko2022epithermal}.
In particular, HSnCT enables the investigation of Bragg edges \cite{tremsin2010braggedge}, which result from Bragg scattering in polycrystalline materials.
To determine the critical Bragg edges from the transmission spectra, HSnCT is often carried out by measuring radiographs in the cold and thermal neutron  \cite{tremsin2009coldthermal, al-falahat2022coldthermal, bakhtiari2020cold} energy range.

Spectral and hyperspectral imaging have proven to be invaluable in other modalities, and several algorithms have been developed to address the complexity of processing and analyzing such high-dimensional data.
A significant milestone in hyperspectral data processing has been the integration of dimensionality reduction techniques like principal component analysis (PCA), independent component analysis (ICA), and non-negative matrix factorization (NMF) that enable lower-dimensional representation of the data for feature extraction, hyperspectral unmixing, and data compression purposes \cite{paul2023othermod, huang2023othermod}.
NMF, in particular, excels in phase unmixing tasks, such as in energy-dispersive X-ray spectroscopy (EDXS), where it leverages a priori knowledge of the sample to improve material identification and separation \cite{chen2024othermod}.

In the context of CT reconstruction for spectral/hyperspectral data, several algorithms have been proposed to enhance the reconstruction quality \cite{gursoy2015othermod}, but the use of dimensionality reduction is relatively limited.
One notable work is the joint reconstruction and spectrum refinement (JoSR) algorithm for photon-counting-detector CT, which uses NMF to reduce the number of parameters for effective spectra estimation \cite{shen2023othermod}.
Additionally, sinogram domain dimensionality reduction using NMF and other techniques has been shown to be effective in reducing reconstruction requirements for multispectral X-ray CT \cite{kheirabadi2017othermod}.
However, the full potential of dimensionality reduction in different spectral and hyperspectral CT applications remains largely unexplored, particularly in the context of HSnCT.

The conventional approach to reconstructing HSnCT data is direct hyperspectral reconstruction (DHR), which involves reconstructing projection data for each wavelength bin separately \cite{daugherty2023hyperrecon}.
The individual reconstructions can be performed using either a basic algorithm like filtered back projection (FBP) \cite{willemink2019fbp, pan2009fbp} or an advanced algorithm like model-based iterative reconstruction (MBIR) \cite{bouman2022mbimaging, liu2014mbir, katsura2012mbir}.
However, even with a fast algorithm like FBP, this approach is very computationally expensive since it requires hundreds of tomographic reconstructions.
Additionally, the reconstructions often suffer from significant noise and artifacts due to the low signal-to-noise ratio (SNR) of each spectral component.

One commonly used approach to HSnCT material decomposition is reconstruction domain material decomposition (RDMD).
In RDMD, hyperspectral reconstructions are first computed using DHR.
The set of linear attenuation coefficients for each material, referred to as the $\mu$-spectrum, is then obtained by computing the vector mean of the corresponding material region within these reconstructions \cite{ametova2021thermal}.
Finally, these $\mu$-spectra are used to decompose the materials.
However, RDMD is time-consuming due to its reliance on DHR, and it is challenging to accurately perform material decomposition using noisy hyperspectral DHR reconstructions.

Alternatively, sinogram domain material decomposition (SDMD) directly performs material decomposition in the sinogram domain \cite{balke2024epithermal, balke2021epithermal}.
Then, only a single reconstruction is performed for each constituent material.
If the $\mu$-spectra are known exactly, then SDMD produces much less noisy reconstructions compared to RDMD and is much faster.
However, obtaining prior knowledge of the exact $\mu$-spectra is difficult in practice.
On the other hand, if the $\mu$-spectra are unknown, then accurate separation of the materials in the sinogram domain is difficult due to both the low SNR and the overlaps of materials in the projections.

\begin{figure}[t!]
\centering
\centerline{\includegraphics[width=0.94\linewidth]{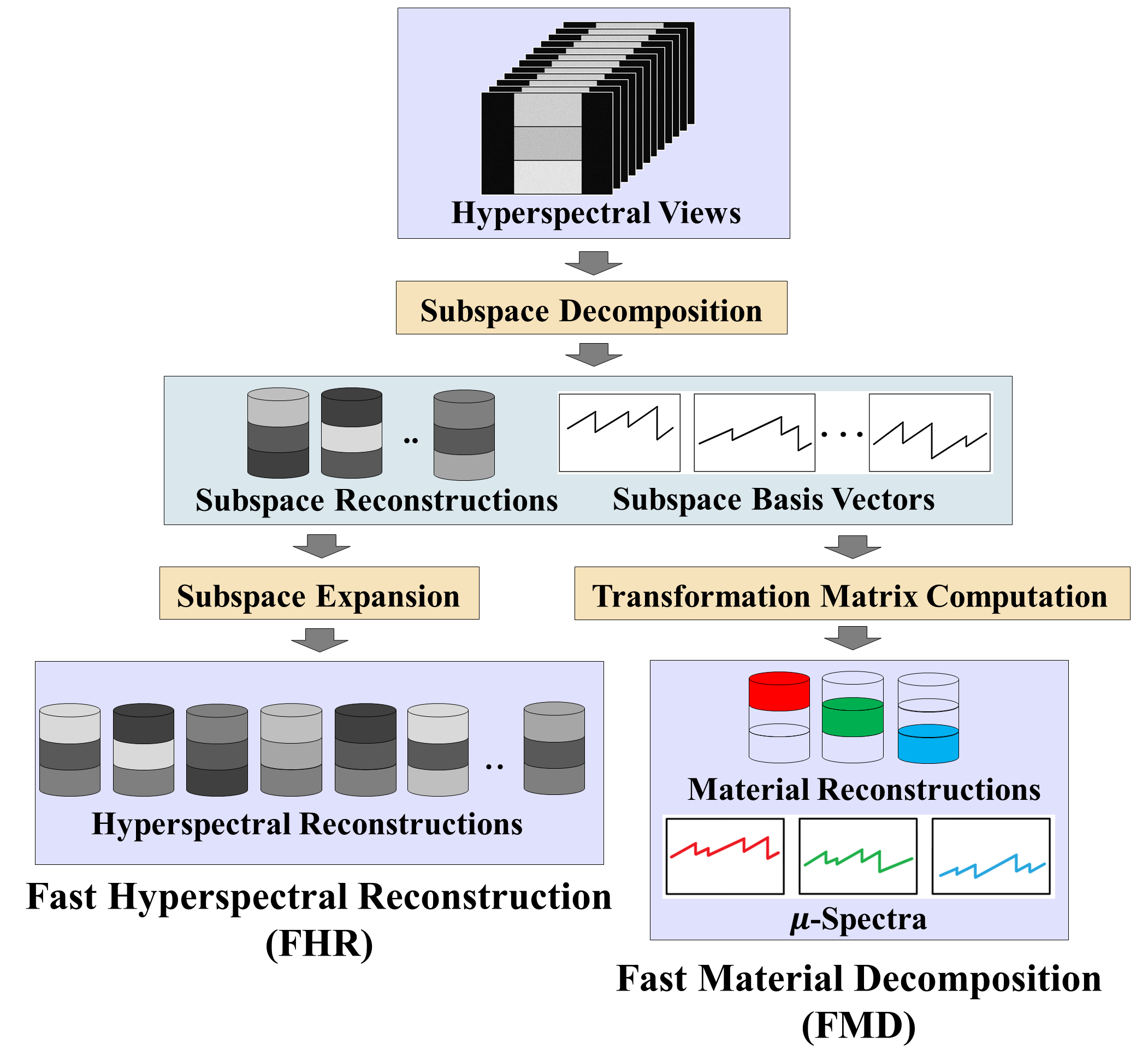}}
\caption{Overview of fast hyperspectral reconstruction (FHR) and fast material decomposition (FMD) algorithms.
Both algorithms use a subspace decomposition procedure to represent the high-dimensional hyperspectral views in a low-dimensional intermediate subspace.
The FHR algorithm then directly performs hyperspectral reconstruction, while the FMD algorithm reconstructs the individual material components.
Both algorithms benefit from increased speed, reduced noise, and reduced artifacts due to the subspace decomposition. 
}
\label{fig:overview}
\end{figure}

In this paper, we present two algorithms for processing HSnCT data.
These algorithms, which we refer to as fast hyperspectral reconstruction (FHR) and fast material decomposition (FMD), are both based on subspace decomposition of the hyperspectral data.
FHR is a method for fast tomographic reconstruction of hyperspectral data that also visibly and quantitatively reduces noise and improves reconstruction quality.
FMD is a related method to perform material decomposition, producing 3D reconstructions of component materials and the associated $\mu$-spectra.
Note that FMD estimates the $\mu$-spectra directly from the data rather than relying on theoretically computed values.
This is because the theoretical values do not account for experiment-dependent physical effects, making them deviate from the true $\mu$-spectra for a particular experiment.
This research extends the work first presented in conference proceedings ~\cite{chowdhury2023amd, chowdhury2025fhr}.

Figure~\ref{fig:overview} illustrates the core components of the FHR and FMD algorithms.
Both algorithms use a subspace decomposition procedure to represent the high-dimensional hyperspectral views in a low-dimensional intermediate subspace and then perform MBIR reconstruction of each subspace component.
The FHR algorithm then directly computes the hyperspectral reconstructions.
Alternatively, the FMD algorithm then volumetrically decomposes the object's materials, i.e., computes the 3D reconstructions of individual materials and their associated $\mu$-spectra.

The use of subspace decomposition in our algorithms serves three purposes:
\begin{itemize}
\item Eliminates significant spectral noise from the data while fitting them into the low-dimensional subspace.
\item Reduces computation time more than 10x due to the reduced number of tomographic reconstructions.
\item Allows for the use of MBIR reconstruction that produces better reconstruction quality from sparse view data.
\end{itemize}

We apply our algorithms to both simulated and measured neutron data and demonstrate that they are substantially faster and yield more accurate results compared to traditional HSnCT methods.

\begin{figure*}[t!]
\centering
\centerline{\includegraphics[width=0.86\linewidth]{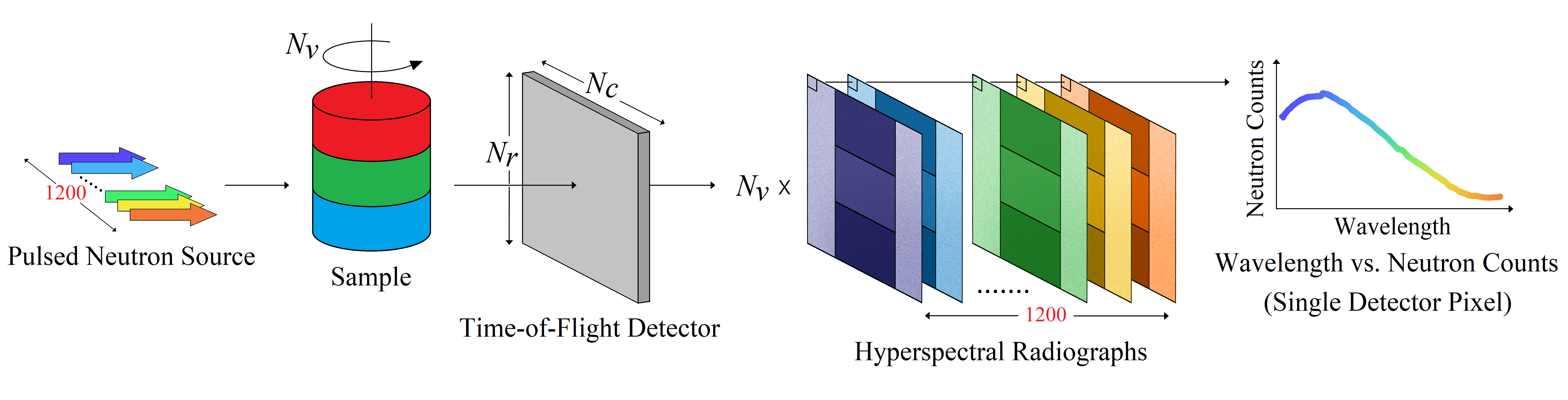}}
\vspace{-0.5cm}
\caption{Illustration of an HSnCT imaging setup at a spallation neutron source.
The arrows on the left represent a pulsed neutron source, which generates a beam of neutrons across a range of wavelengths.
The hyperspectral beam travels through the sample to the $N_r \times N_c$ time-of-flight (TOF) imaging detector, which records the number of neutrons as a function of time.
Based on the time of arrival, these detector counts are resolved into $N_k=1200$ wavelength-specific radiographs.
The plot on the right shows neutron counts across the wavelength bins for a single detector pixel.
For tomographic reconstruction, wavelength-resolved data are collected for $N_v$ orientations of the sample.}
\label{fig:img_sys}
\end{figure*}

\section{The HSnCT Imaging System}
\label{sec:Imaging_System}

Figure~\ref{fig:img_sys} illustrates a simple HSnCT imaging system that is used to collect wavelength-resolved hyperspectral data from multiple views of the sample.
HSnCT has commonly been conducted using a pulsed neutron source that relies on the spallation of neutrons \cite{strobl2009sns, kockelmann2018sns, nelson2018sns}.
The pulse of neutrons passes through the sample and is detected by a 2D time-of-flight (TOF) imaging array \cite{tremsin2012tof, watanabe2017tof, Kiyanagi2005tof}.
The TOF detector counts the number of neutrons at each pixel and for each time interval bin.
These time interval bins then correspond to each neutron's velocity or wavelength.
The specific relationship between the neutron TOF, $\Delta t$, and its wavelength, $\lambda$, is given by
\begin{equation} \label{eq:wave_tof}
    \lambda = \frac{h}{m_n} \frac{\Delta t}{L} \ ,
\end{equation}
where $h$ is Planck's constant, $m_n$ is the neutron mass, and $L$ is the distance between the source and the detector.

In order to describe our method, we introduce the following notation:
\begin{itemize}
    \item $N_r$ is the number of detector rows
    \item $N_c$ is the number of detector columns
    \item $N_k$ is the number of wavelength bins
    \item $N_v$ is the number of tomographic views
    \item $N_m$ is the number of materials in the sample
    \item $N_p=N_v \times N_r \times N_c$ is the number of projections
    \item $N_x=N_r \times N_c \times N_c$ is the number of voxels
\end{itemize}

The output of the TOF detector is a hyperspectral neutron radiograph \cite{josic2010energy} in the form of an $N_r \times N_c \times N_k$ array.
For a typical HSnCT system with $N_r=N_c=512$ and $N_k=1200$, a single hyperspectral radiograph will have $3.146 \times 10^8$ data points, or equivalently, 300 megapixels of data.

In order to perform a tomographic scan, the object is rotated to $N_v$ view orientations \cite{yang2023orientation, tang2024orientation, haque2013orientation, dabravolski2014orientation} and at each orientation, a hyperspectral radiograph is measured.
This results in a tomographic sinogram of neutron counts with the form $y_{v,r,c,k}$ where $v,r,c,k$ are the discrete view, row, column, and wavelength indices.
In addition, a single hyperspectral radiograph is measured with the object removed.
This is known as the ``open-beam'' and is denoted by $y^o_{r,c,k}$.
From these, we can compute the hyperspectral projection views, $p$, using the relationship \cite{Vontobel2006projdensity} given by
\begin{equation}
\label{eq:neglogattenuation}
p_{v,r,c,k} = -\log \left( \frac {y_{v,r,c,k}} {y_{r,c,k}^o} \right) \ .
\end{equation}
$p_{v,r,c,k}$ corresponds to the line integral of the linear attenuation coefficient along the projection at detector location $(r,c)$ for wavelength bin $k$ and orientation $v$.
Note that in practice, various corrections must be made in the calculation of \eqref{eq:neglogattenuation} in order to account for effects such as scatter, detector bias, and count rate limitations. 
Details are given in Appendix~\ref{appendix:off_corr}.

In order to formulate a forward model for the HSnCT system, let $x^m \in \mathbb{R}^{N_x \times N_m}$ denote the sample where each column of $x^m$ represents the 3D volume fraction for a single material.
Furthermore, let $D^m \in \mathbb{R}^{N_k \times N_m}$ be a dictionary of spectral responses where each column is the vector of linear attenuation coefficients as a function of wavelength for a single material.
Also, let $p \in \mathbb{R}^{N_p \times N_k}$ be the set of hyperspectral projection views.

Using these definitions, we assume that the forward model has the form
\begin{equation} \label{eq:porj_matd_rel}
    p = (Ax^m) (D^m)^\top + w \ ,    
\end{equation}
where $A\in \mathbb{R}^{N_p\times N_x}$ is the linear projection operator, 
$w$ represents additive noise,
and $\top$ denotes transpose.

\begin{figure*}[t!]
\centering
\centerline{\includegraphics[width=0.85\linewidth]{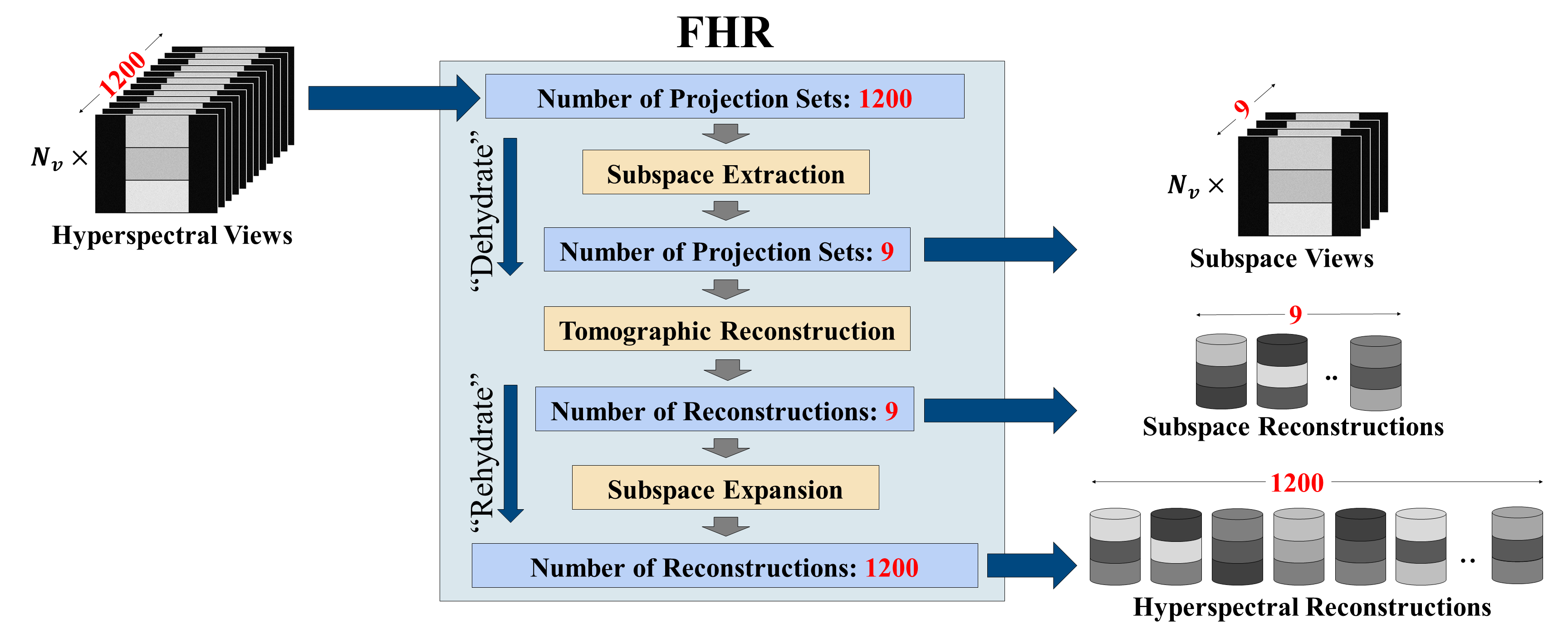}}
\vspace{-0.4cm}
\caption{Illustration of the FHR algorithm with sample inputs (hyperspectral views) and outputs (hyperspectral reconstructions).
FHR first transforms the $N_k=1200$ dimensional hyperspectral views into $N_s=9$ dimensional subspace views.
Then it performs tomographic reconstruction using MBIR to produce 9 subspace reconstructions.
Finally, it expands the 9 subspace reconstructions into 1200 hyperspectral reconstructions using the basis vectors of the subspace.}
\label{fig:FHR_pipeline}
\end{figure*}

\section{Fast Hyperspectral Reconstruction (FHR)}
\label{sec:Fast_Hyper_Recon}

Figure~\ref{fig:FHR_pipeline} illustrates the three steps in FHR consisting of subspace extraction, MBIR reconstruction, and subspace expansion.
FHR reduces the spectral dimension during subspace extraction, then performs volumetric reconstruction in this lower-dimensional space, and finally restores the spectral dimension using subspace expansion.
We can think of the algorithm as ``dehydrating'' the data in the sinogram domain and then ``rehydrating'' in the reconstruction domain.

\begin{figure}[b!]
\centering
\includegraphics[width=0.85\linewidth]{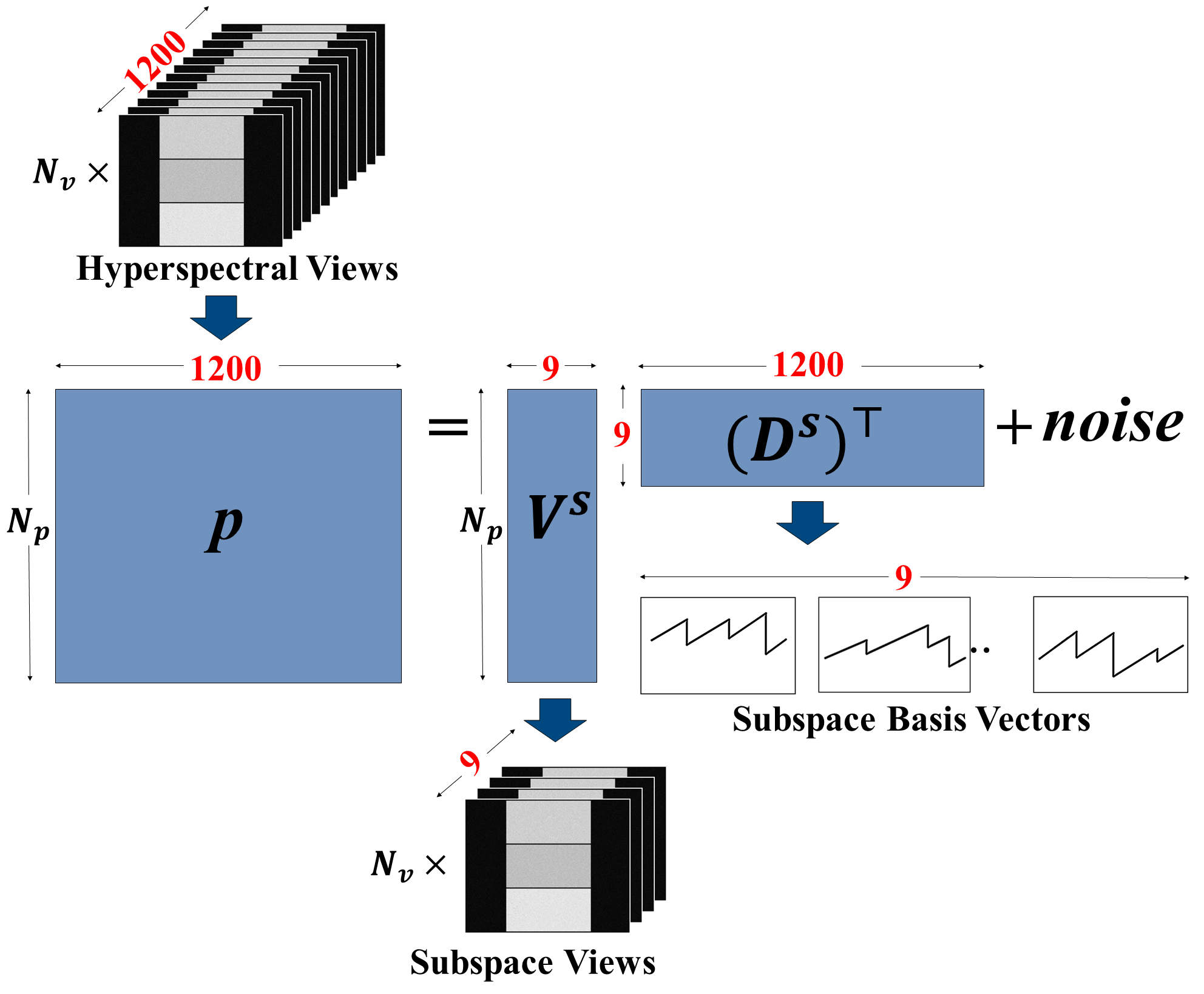}
\caption{Subspace extraction:
FHR employs NMF to decompose the $N_k=1200$ dimensional hyperspectral views ($p$) into $N_s=9$ dimensional subspace views ($V^s$) along with the corresponding subspace basis vectors ($D^s$).
This approach effectively reduces the data dimensions and also significantly reduces spectral noise.}
\label{fig:sub_ext}
\end{figure}

Figure~\ref{fig:sub_ext} illustrates the subspace extraction step of FHR in which the high-dimensional hyperspectral views $p \in \mathbb{R}^{N_p \times N_k}$ are decomposed into low-dimensional subspace views $V^s \in \mathbb{R}^{N_p \times N_s}$ and a corresponding dictionary of subspace basis vectors $D^s \in \mathbb{R}^{N_k \times N_s}$.
Here, $N_s$ is the dimension of the intermediate subspace, and typically $N_m \leq N_s<<N_k$.
The decomposition can then be obtained by solving the non-negative matrix factorization (NMF) \cite{pauca2006nmf, lee2000nmf, zhang2022nmf} problem given by
\begin{equation} \label{eq:sub_extract_solve}
    (V^s, D^s) = \argmin_{(V\geq 0, D\geq 0)} \{ \Vert p - V D^\top \Vert_{F}^2 \} \ .
\end{equation}
NMF was implemented with non-negative double singular value decomposition (NNDSVD) initialization and coordinate descent solver that uses fast hierarchical alternating least squares (Fast HALS).
For the software implementation, we used the scikit-learn Python package \cite{scikit-learn}.

Since $N_s<<N_k$, operations performed in the subspace domain are much faster compared to those in the hyperspectral domain.
Also, the residual difference from the decomposition $\epsilon = p-V^s (D^s)^\top$ primarily consists of spectral noise, which is effectively removed from the data.

The choice of $N_s$ generally depends on the number of materials in the sample.
We choose $N_s = \lceil \beta N_m \rceil$, where $\beta \geq 1$ is a user-selectable parameter.
Ideally, one might expect that $\beta=1$ would be the best choice.
However, in Section~\ref{ssec:subspace_dim}, we demonstrate that it is better to pick $N_s$ to be larger than the number of materials.
When $N_s=N_m$, then the intermediate subspace has exactly the required dimension under ideal conditions. However, in practice, the signal can fall outside the selected subspace due to noise-induced error in the subspace estimation or nonlinearities in the system that increase the dimensionality of the signal. We note that system nonlinearities can arise from physical effects such as neutron scattering, partial volume effects, energy resolution effects, and detector physics.
Consequently, in this research, we will use $\beta=3$.

\begin{figure}[t!]
\centering
\includegraphics[width=0.85\linewidth]{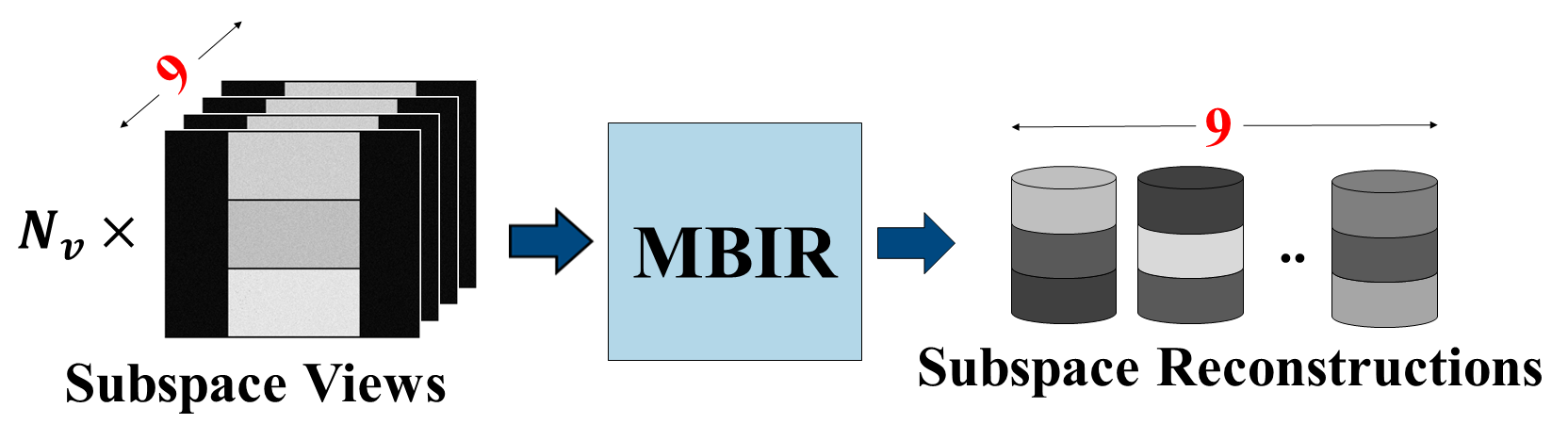}
\caption{Tomographic reconstruction:
FHR computes $N_s=9$ reconstructions ($x^s$) from the extracted 9 sets of subspace views ($V^s$) using MBIR.
This allows the algorithm to transition from the sinogram domain to the spatial domain.}
\label{fig:recon}
\end{figure}

Figure~\ref{fig:recon} illustrates the tomographic reconstruction step of FHR that computes $N_s$ reconstructions within the subspace $x^s \in \mathbb{R}^{N_x \times N_s}$ using the MBIR algorithm \cite{bouman2022mbimaging, liu2014mbir, katsura2012mbir}.
For each subspace index $j=1, 2, \ldots, N_s$, MBIR solves the optimization problem given by
\begin{equation} \label{eq:recon}
    x_j^s = \argmin_{x_j} \left \{ \frac{1}{2 \sigma_v^2} \Vert V_j^s - Ax_j \Vert^2 + h(x_j) \right \} \ ,
\end{equation}
where $A$ is the linear projection operator from a rasterized volume to a rasterized sinogram, $V_j^s$ is the $j^{th}$ column of $V^s$, and $h(x_j)$ is the Q-Generalized Gaussian Markov random field (qGGMRF) prior model \cite{wang2011qggmrf, kisner2012qggmrf, pellizzari2017qggmrf}.
$\sigma_v$ is the assumed noise standard deviation.
Since we perform reconstruction in the subspace domain, the traditional noise model does not apply directly; therefore, we use a simple model that uses a constant $\sigma_v$ across pixels. For the software implementation, we used the SVMBIR Python package \cite{wang2016svmbir, svmbir-code}.

\setcounter{figure}{6}
\begin{figure*}[b!]
\centering
\centerline{\includegraphics[width=0.85\linewidth]{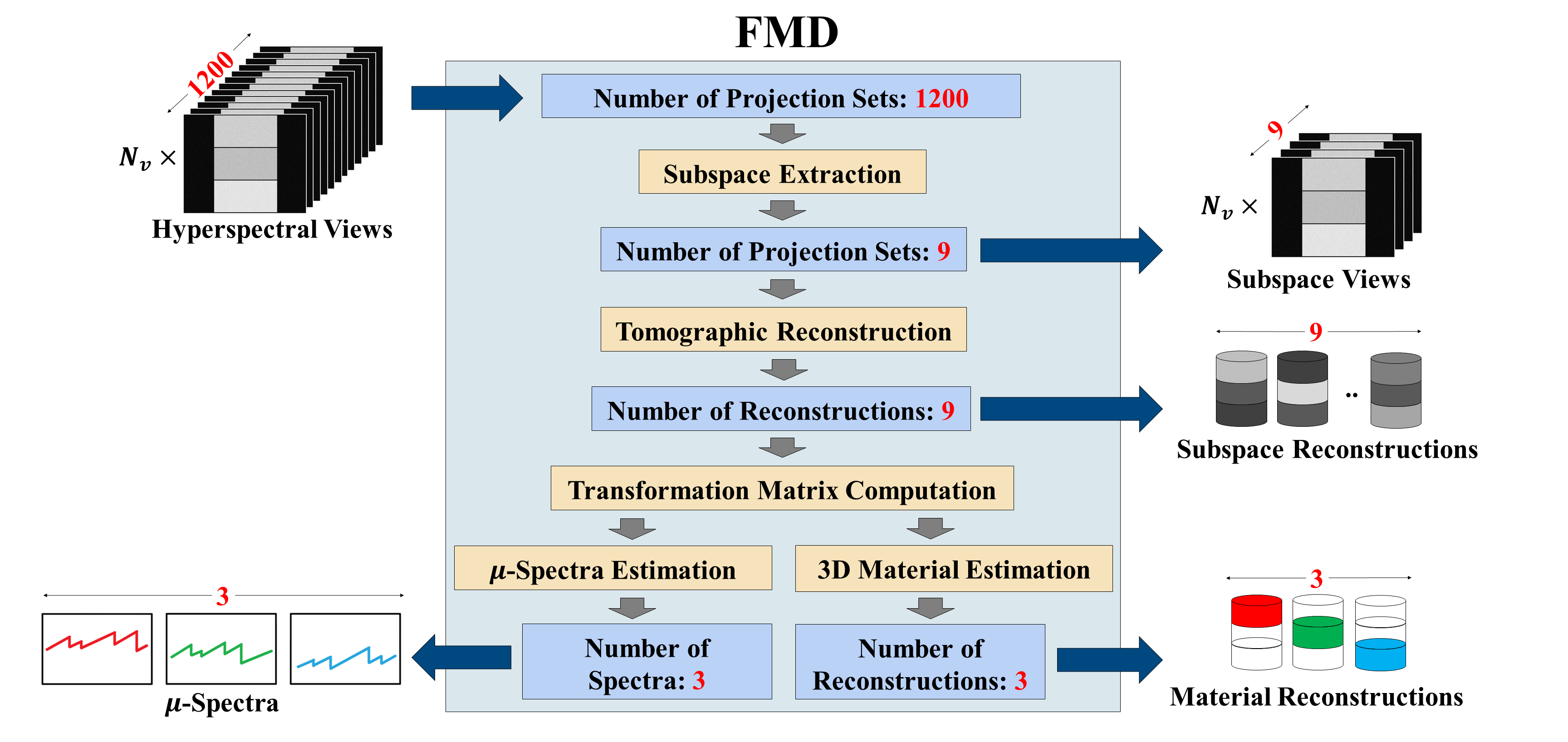}}
\vspace{-0.4cm}
\caption{Illustration of the FMD algorithm with sample inputs (hyperspectral views) and outputs (reconstructed materials, $\mu$-spectra).
Similar to FHR, FMD first performs the subspace decomposition procedure.
Then, a transformation matrix is calculated from the $N_s=9$ subspace reconstructions.
The algorithm uses this transformation matrix to transform the 9 subspace reconstructions into $N_m=3$ material reconstructions and the 9 subspace basis vectors into 3 material $\mu$-spectra.}
\label{fig:FMD_pipeline}
\end{figure*}

\setcounter{figure}{5}
\begin{figure}[t!]
\centering
\includegraphics[width=0.85\linewidth]{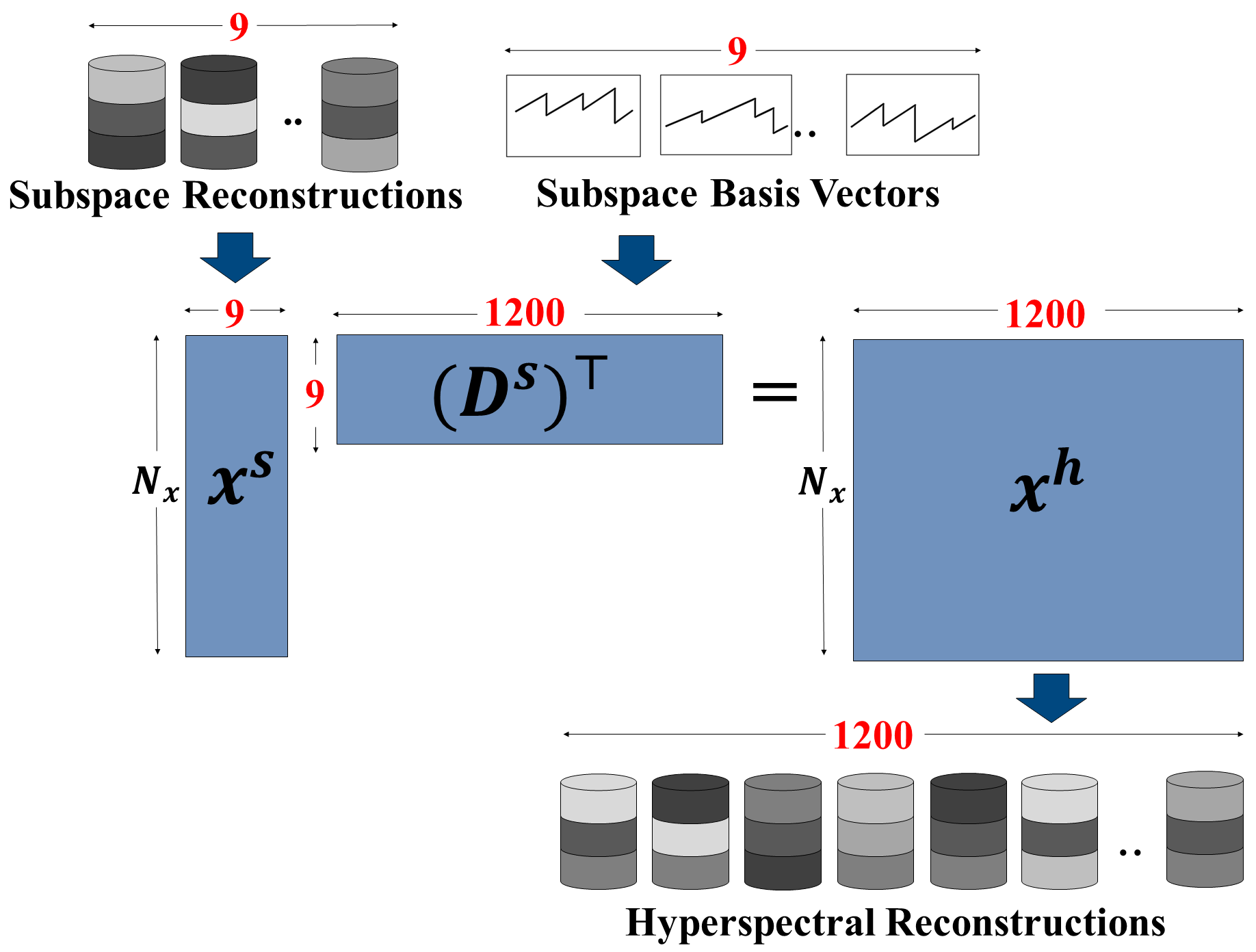}
\caption{Subspace expansion:
FHR expands the $N_s=9$ subspace reconstructions ($x^s$) into $N_k=1200$ hyperspectral reconstructions ($x^h$) using the subspace basis vectors ($D^s$).}
\label{fig:sub_expansion}
\end{figure}
\setcounter{figure}{7}

\begin{algorithm}[t!]
    \KwInput{$p, N_s$}
    \KwOutput{$x^h$}
    
    \vspace{0.3cm}
    \tcp{Subspace Extraction}
    $(V^s, D^s) \gets \argmin\limits_{(V\geq 0, D\geq 0)} \{ \Vert p - V D^\top \Vert_{F}^2$\}
    
    \vspace{0.3cm}
    \tcp{Tomographic Reconstruction}
    \For{$j \gets 1$ \KwTo $N_s$}{
        $x_j^s \gets \argmin\limits_{x_j} \left \{ \frac{1}{2 \sigma_v^2} \Vert V_j^s - Ax_j \Vert^2 + h(x_j) \right \}$
    }

    \vspace{0.3cm}
    \tcp{Subspace Expansion}
    $x^h \gets x^s(D^s)^\top$
    
    \vspace{0.3cm}
    \Return{$x^h$}
\caption{Fast Hyperspectral Reconstruction}
\label{alg:FHR}
\end{algorithm}

MBIR is capable of producing high-quality reconstructions from sparse-view and low-SNR measurements.
As a result, it can effectively reduce the data acquisition time for HSnCT.
On the downside, MBIR reconstruction tends to be much slower than FBP reconstruction.
However, the increased reconstruction time for MBIR is of much less concern since FHR only requires the reconstruction of $N_s$ 3D volumes rather than the $N_k$ 3D volumes required for DHR.

Figure~\ref{fig:sub_expansion} illustrates the subspace expansion step of FHR in which the algorithm expands the subspace reconstructions into hyperspectral reconstructions using $D^s$. 
Since $D^s$ maps each voxel from subspace to hyperspectral coordinates, the hyperspectral reconstructions $x^h \in \mathbb{R}^{N_x \times N_k}$ can be calculated using the relationship given by
\begin{equation} \label{eq:sub_expansion}
    x^h = x^s(D^s)^\top.
\end{equation}

The entire procedure for FHR is summarized in Algorithm \ref{alg:FHR}.

\section{Fast Material Decomposition (FMD)}
\label{sec:Fast_Material_Decomposition}

\begin{figure}[t!]
\centering
\includegraphics[width=0.75\linewidth]{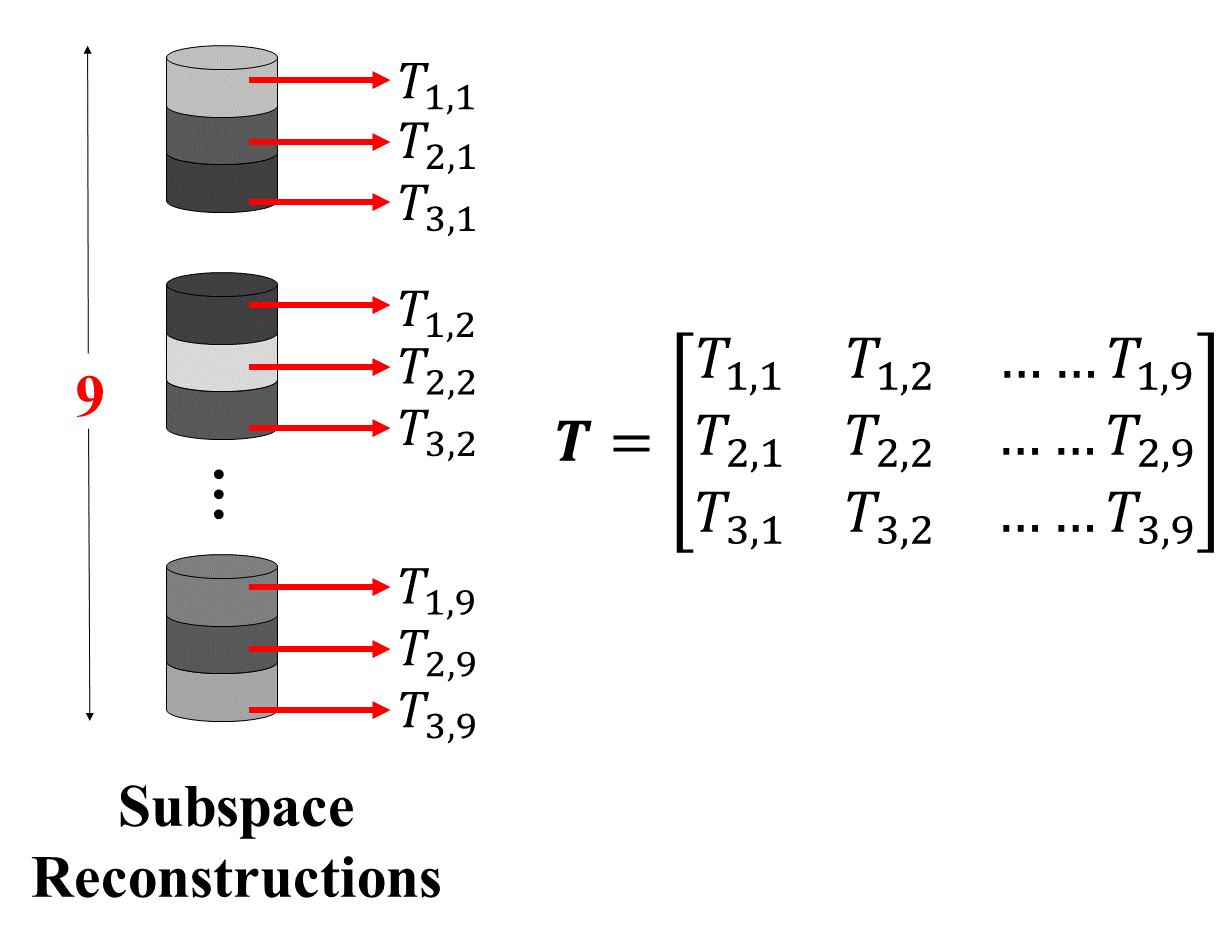}
\vspace{-0.2cm}
\caption{Transformation matrix computation:
To go from $N_s=9$ dimensional subspace to $N_m=3$ materials, FMD computes a $3 \times 9$ transformation matrix $T$ from the subspace reconstructions ($x^s$).
This transformation matrix serves as a bridge between the subspace domain and the material domain.}
\label{fig:trans_mat}
\end{figure}

\begin{figure}[b!]
\centering
\includegraphics[width=0.75\linewidth]{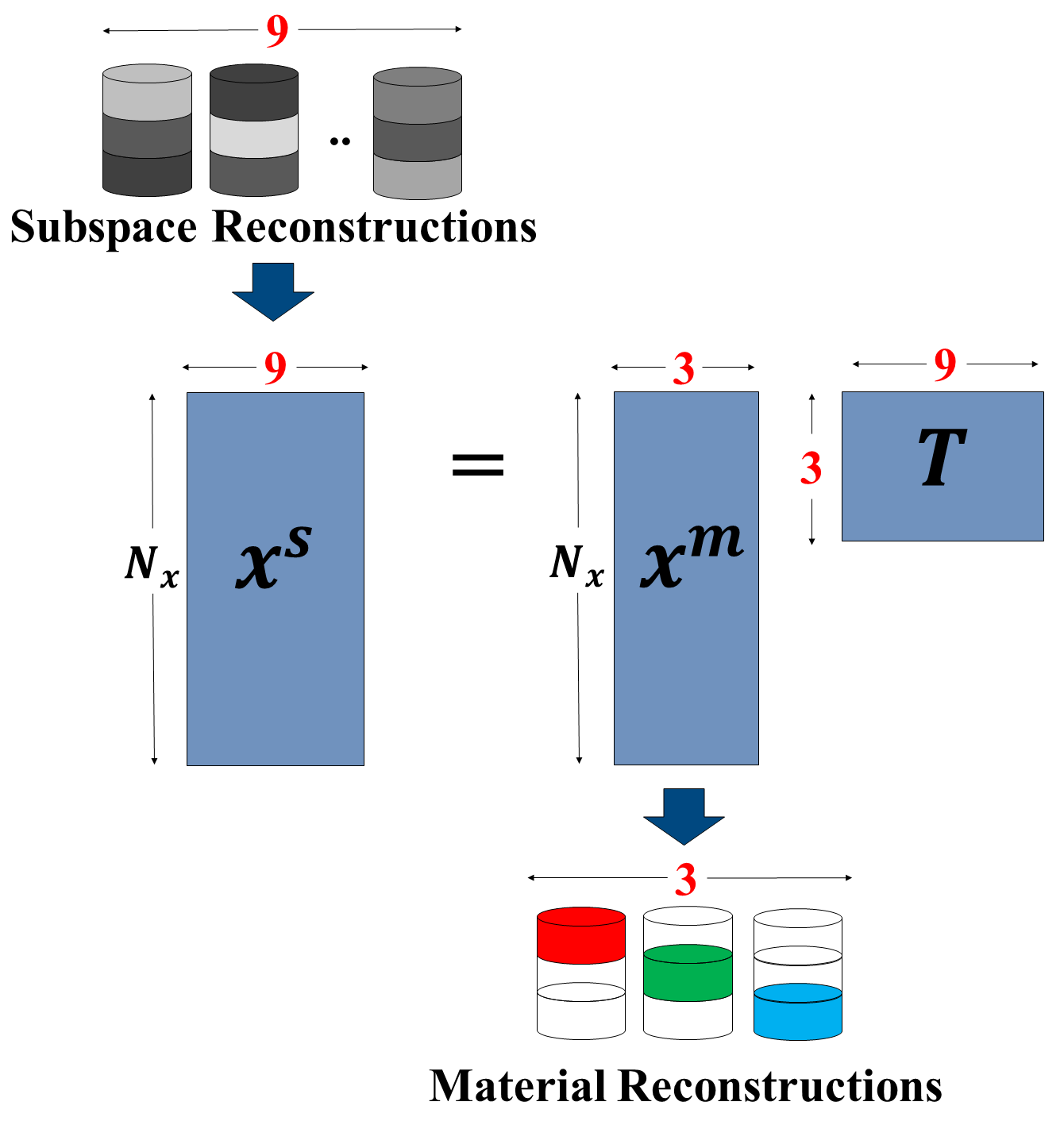}
\caption{3D material estimation: 
FMD transforms the $N_s=9$ subspace reconstructions ($x^s$) into $N_m=3$ material reconstructions ($x^m$) using the transformation matrix ($T$).
These spatially separated material volumes are the intended outputs of the FMD algorithm.}
\label{fig:mat_recon}
\end{figure}

Next, we illustrate how our approach can be used to reconstruct individual materials/crystallographic phases and their associated attenuation spectra. 
Figure~\ref{fig:FMD_pipeline} illustrates how the FMD algorithm first performs subspace extraction and reconstruction and then calculates the material transformation matrix that converts the subspace reconstructions and basis vectors into reconstructions of the individual materials along with the associated $\mu$-spectra for each material.
We note that the first two steps - subspace extraction and reconstruction are identical to the steps used in FHR.
We will describe the remaining steps required to perform FMD. 

Figure~\ref{fig:trans_mat} illustrates the material transformation matrix computation step of FMD in which an $N_m \times N_s$ matrix $T$ is computed to transition from the intermediate subspace to physically meaningful material space.
This matrix is computed either by using the homogeneous material regions provided by the users (semi-supervised mode) or by estimating these homogeneous regions using a clustering procedure (unsupervised mode).

For the unsupervised mode, FMD implements a Gaussian mixture model-based clustering procedure \cite{bouman2022mbimaging, reynolds2009gaussian, xuan2001gaussian} on $x^s$ and segments the material regions based on the estimated model parameters.
GMCluster~\cite{gmcluster} Python package is used for the implementation of this clustering.
To refine the segmentation, a morphological closing operation is performed, which fills any small holes within the regions.
This is followed by a morphological erosion operation that removes the region borders, as the bordering voxels may contain overlapping materials.
This ensures that the final homogeneous material regions are well-separated.
Both morphological closing and erosion are applied using an $N_q \times N_q$ neighborhood.

\begin{algorithm}[t!]
    \KwInput{$p, N_s, N_m, M(optional)$}
    \KwOutput{$x^m, D^m$}
    
    \vspace{0.3cm}
    \tcp{Subspace Extraction}
    $(V^s, D^s) \gets \argmin\limits_{(V\geq 0, D\geq 0)} \{ \Vert p - V D^\top \Vert_{F}^2$\}
    
    \vspace{0.3cm}
    \tcp{Tomographic Reconstruction}
    \For{$j \gets 1$ \KwTo $N_s$}{
        $x_j^s \gets \argmin\limits_{x_j} \left \{ \frac{1}{2 \sigma_v^2} \Vert V_j^s - Ax_j \Vert^2 + h(x_j) \right \}$
    }

    \vspace{0.3cm}
    \tcp{Homogeneous Region Segmentation}
    \If{$M = \emptyset$}{
        $M \gets \mbox{cluster\_segment}(x^s)$
    }
    
    \vspace{0.3cm}
    \tcp{Transformation Matrix Computation}
    \For{$i \gets 1$ \KwTo $N_m$}{
    \For{$j \gets 1$ \KwTo $N_s$}{
        $T_{i,j} \gets \frac{1}{\vert M_i \vert} \mathlarger{\sum}\limits_{n \in M_i} x_{n, j}^s$
        }
        }
    
    \vspace{0.3cm}
    \tcp{3D Material Estimation}
    $x^m \gets \argmin\limits_{x\geq 0} \{ \Vert x^s-x T \Vert_{F}^2 \}$

    \vspace{0.3cm}
    \tcp{$\mu$-Spectra Estimation}
    $D^m \gets D^s T^\top$

    \vspace{0.3cm}
    \Return{$x^m, D^m$}
\caption{Fast Material Decomposition}
\label{alg:FMD}
\end{algorithm}

Once the homogeneous material regions are identified, the vector mean for each region in $x^s$ is computed to populate each row of $T$.
The computation of each element $T_{i,j}$ can be expressed as
\begin{equation} \label{eq:trans_mat}
    T_{i,j} = \frac{1}{\vert M_i \vert} \sum_{n \in M_i} x_{n, j}^s \ ,
\end{equation}
where $M_i$ is the set of voxel indices from $i^{th}$ material region and $\vert M_i \vert$ is the number of voxels in that region.

\begin{figure}[t!]
\centering
\includegraphics[width=0.88\linewidth]{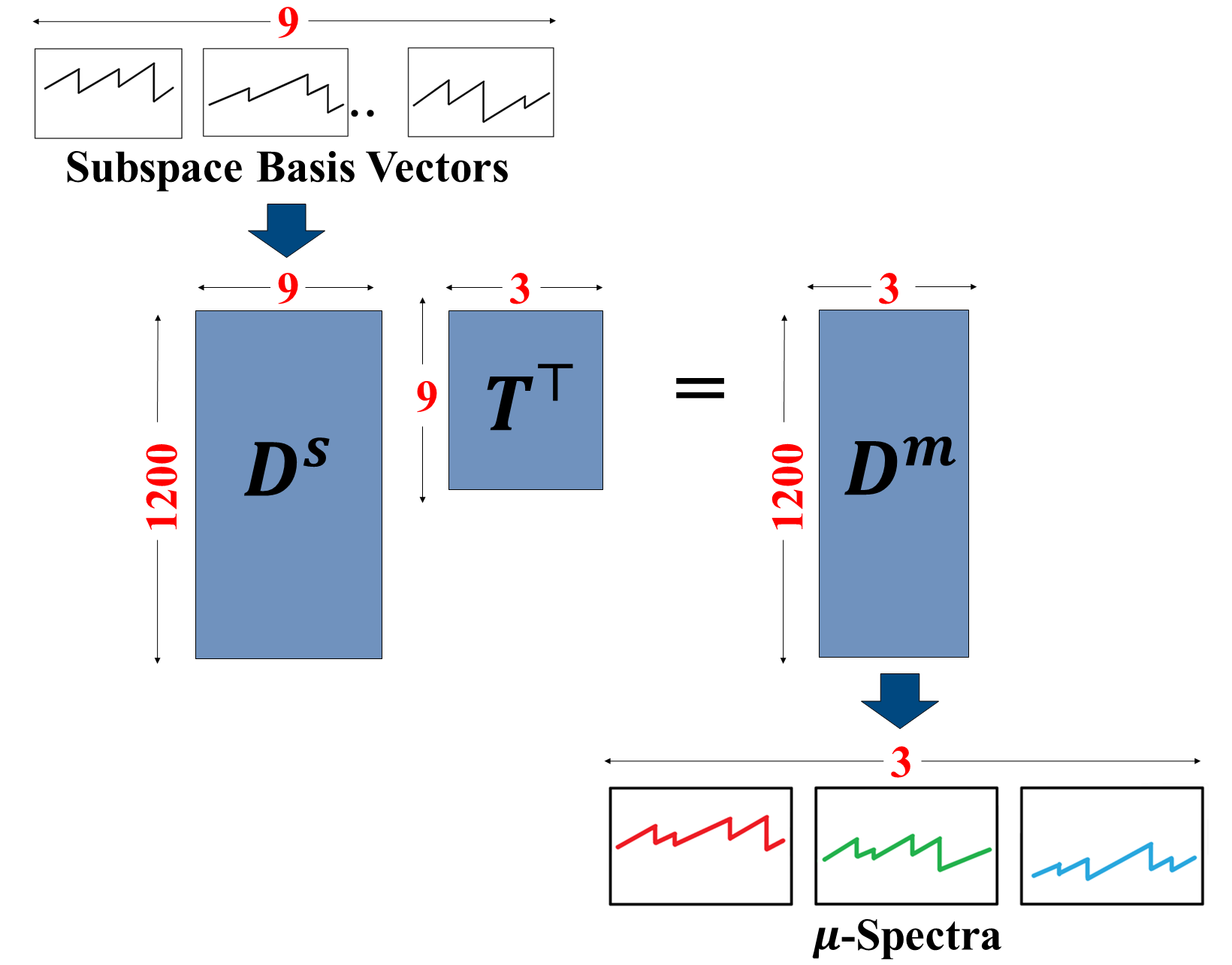}
\caption{$\mu$-spectra estimation: 
FMD uses the transformation matrix ($T$) again to transform the $N_s=9$ subspace basis vectors ($D^s$) into $N_m=3$ material $\mu$-spectra ($D^m$).
Each spectrum retains Bragg edges that are unique to the associated material.}
\label{fig:mat_basis_est}
\end{figure}

Figure~\ref{fig:mat_recon} illustrates the 3D material estimation step of FMD, which transforms the subspace reconstructions into material reconstructions using the transformation matrix $T$.
As each subspace reconstruction can be represented as a weighted sum of all the material reconstructions, and $T$ contains these weights, we can compute the material reconstructions $x^m \in \mathbb{R}^{N_x \times N_m}$ by solving the optimization problem given by
\begin{equation} \label{eq:mat_recon_nmf}
    x^m = \argmin_{x\geq 0} \{ \Vert x^s-x T \Vert_{F}^2 \} \ .
\end{equation}

Figure~\ref{fig:mat_basis_est} illustrates the $\mu$-spectra estimation step of FMD in which the $\mu$-spectra dictionary $D^m$ is computed from the subspace basis matrix $D^s$ using the relationship (Appendix~\ref{appendix:rel_sub_mat_basis}) given by
\begin{equation}\label{eq:sub_mat_rel_basis}
    D^m = D^s T^\top  \ .
\end{equation}

The entire procedure for FMD is summarized in Algorithm \ref{alg:FMD}.

\section{Results}
\label{sec:results}

In order to demonstrate the value of the FHR and FMD algorithms, we applied them to both simulated and measured HSnCT data, and we compared our results to those of the conventional approaches.

\subsection{Data and Methods}
\label{ssec:data}

\begin{figure}[t!]
\begin{minipage}[a]{0.42\linewidth}
\centering
\includegraphics[width=3.5cm]{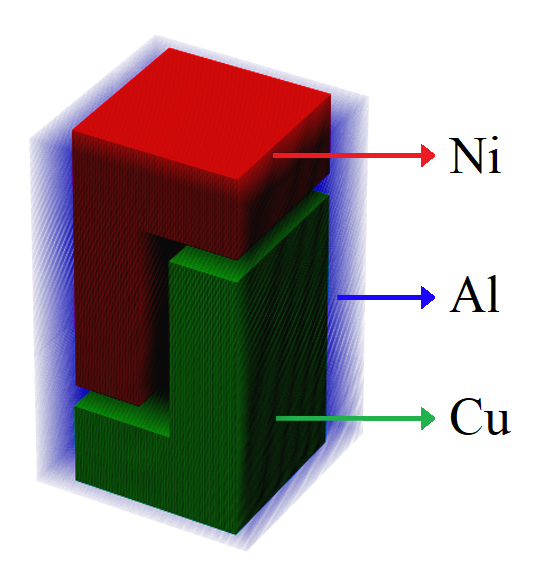}
\centerline{(a)}\medskip  
\end{minipage}
\hfill
\begin{minipage}[a]{0.42\linewidth}
\centering
\includegraphics[width=3.2cm]{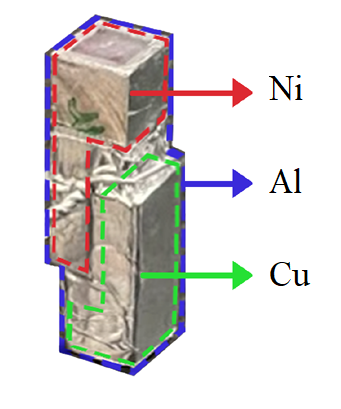}
\centerline{(b)}\medskip
\end{minipage}
\vspace{-0.1in}
\caption{Images of (a) simulated phantom and (b) scanned physical phantom.
The simulated phantom provides a rough approximation of the physical phantom, with both phantoms containing powdered Ni and powdered Cu in Al structures.}
\label{fig:sim_actual_samp}
\end{figure}

The simulated HSnCT data was generated using the forward model described in section \ref{sec:Imaging_System}.
First, we designed a synthetic phantom $x^m$ that contains 3 distinct, non-overlapping materials: nickel (Ni), copper (Cu), and aluminum (Al).
The synthetic phantom loosely approximates the physical phantom, as seen from Figure~\ref{fig:sim_actual_samp}.
Next, we generated a dictionary of realistic $\mu$-spectra $D^m$ using the Bragg-edge modeling (BEM) library \cite{lin2018bem}, which covers a wavelength range of 1.5 to 4.5 \AA.
The material densities for $\mu$-spectra generation were chosen assuming Ni and Cu in powdered form and Al in solid form.
This was done to be consistent with the measured data described below.
Using the simulated $x^m$ and $D^m$, we computed the hyperspectral projection views $p$ from equation~\ref{eq:porj_matd_rel}.
We also simulated hyperspectral blank scans $y^o$ with realistic neutron dosage rates (peak dosage rate: 500).
Then we computed the hyperspectral neutron counts $y$ from the relationship given by:
\begin{equation}
\label{eq:neutroncounts}
y_{v,r,c,k} = y_{r,c,k}^o e^{-p_{v,r,c,k}} \ .
\end{equation}
Finally, we introduced realistic noise characteristics by sampling both $y$ and $y^o$ from the Poisson distribution.

\begin{figure*}[t!]
\begin{minipage}[a]{0.5\linewidth}
\centering
\includegraphics[width=8.6cm]{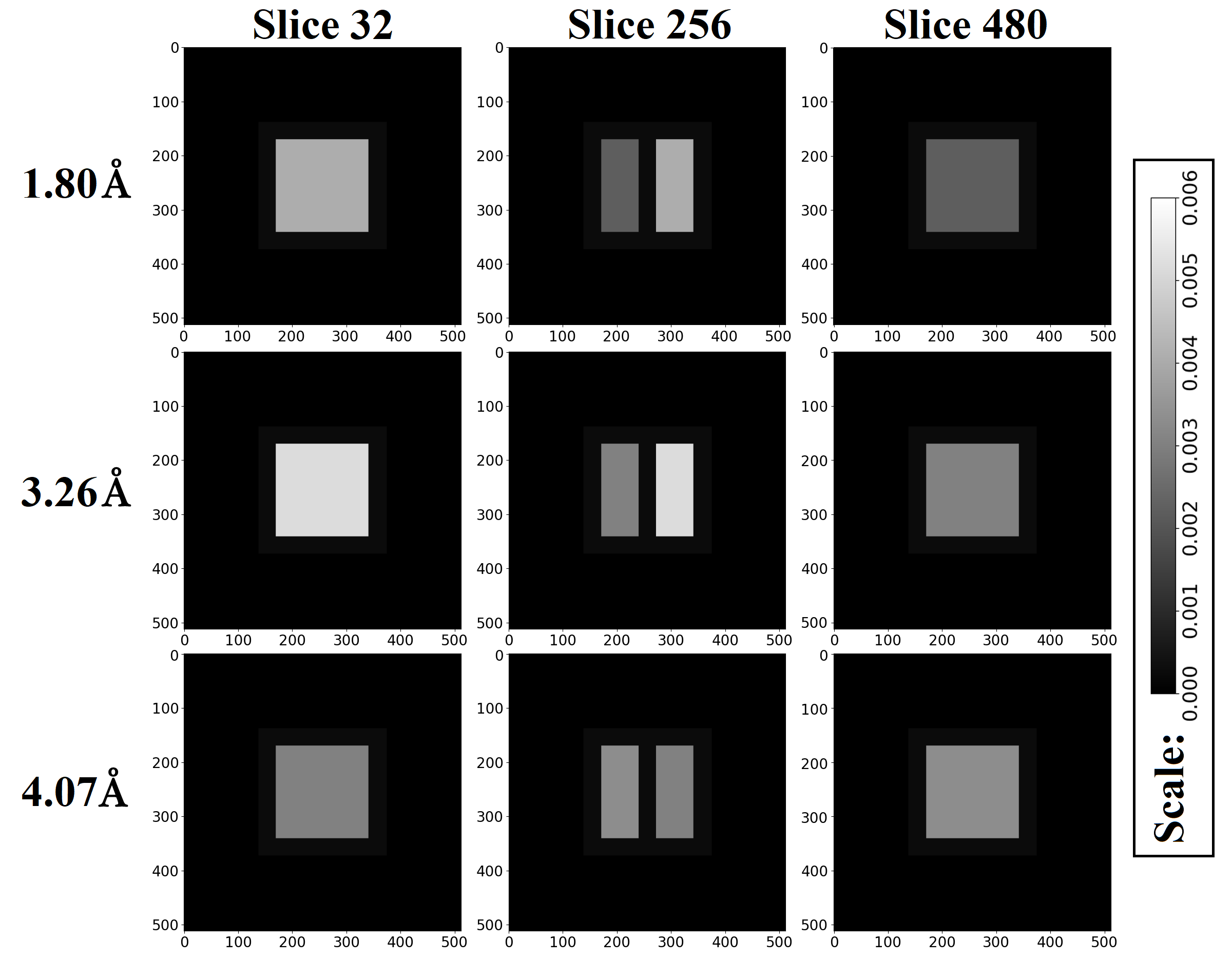}
\centerline{(a) Ground Truth}\medskip  
\end{minipage}
\hfill
\begin{minipage}[a]{0.5\linewidth}
\centering
\includegraphics[width=8.6cm]{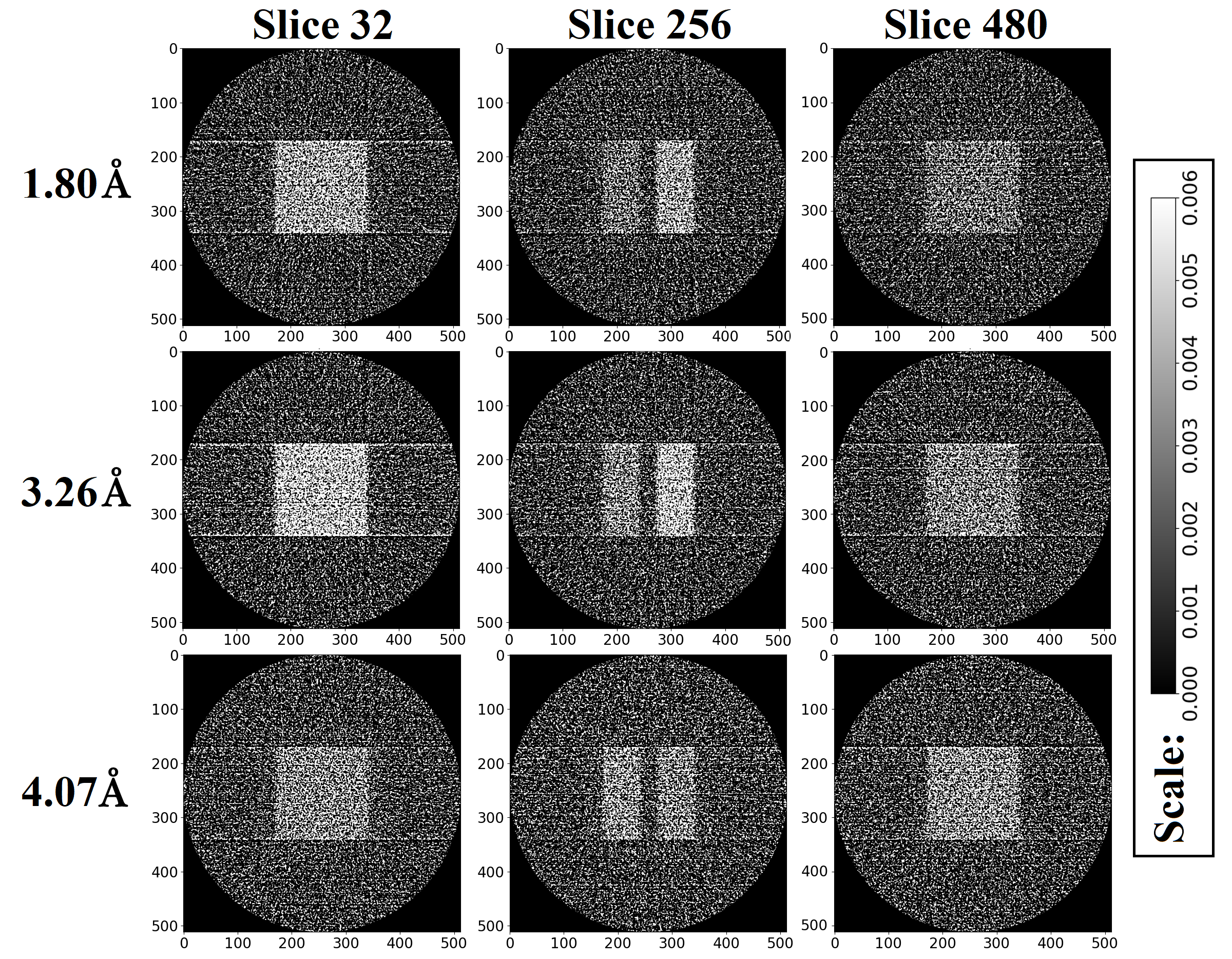}
\centerline{(b) Baseline Method (DHR)}\medskip  
\end{minipage}
\hfill
\begin{minipage}[a]{0.5\linewidth}
\centering
\includegraphics[width=8.6cm]{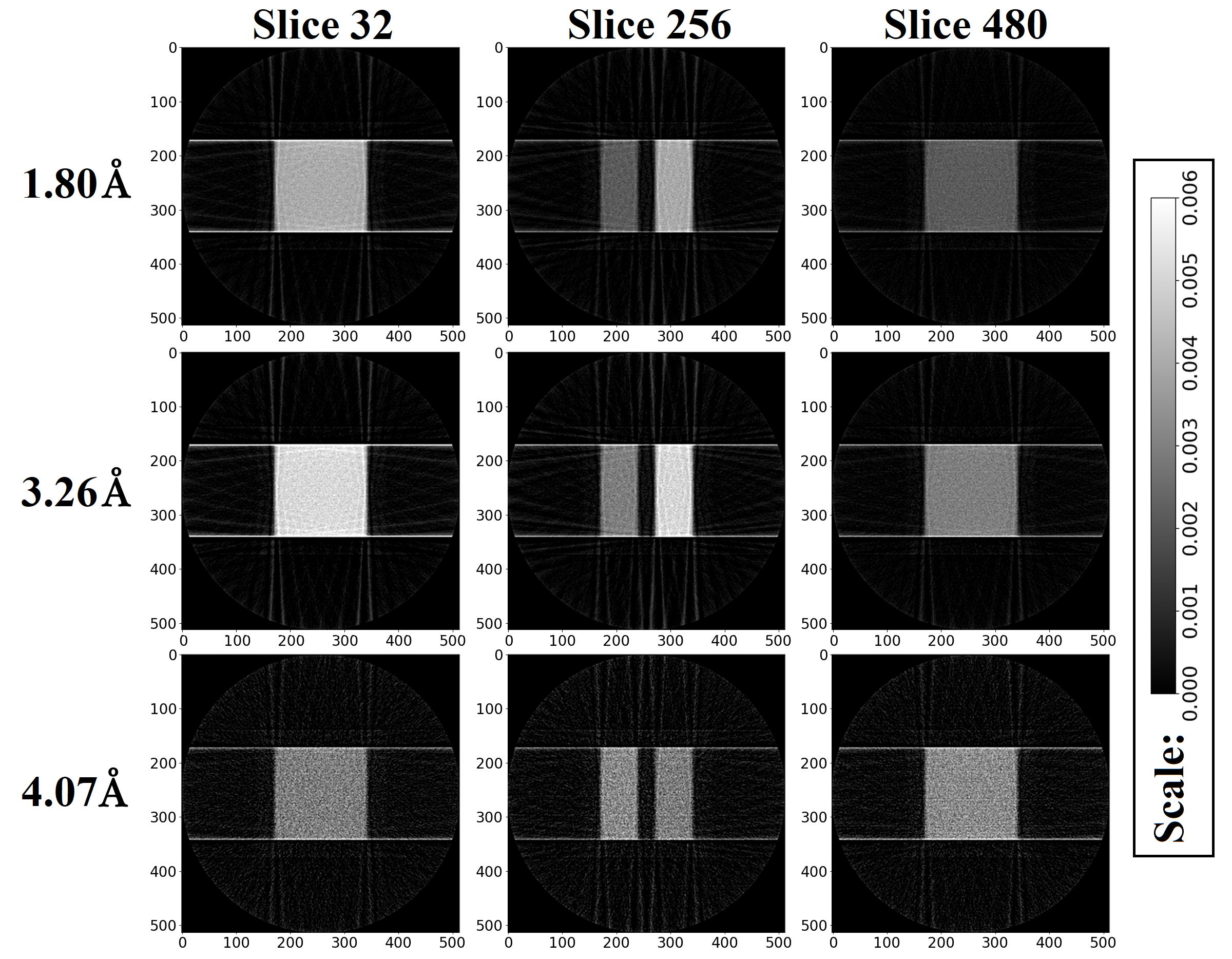}
\centerline{(c) FHR with FBP}\medskip  
\end{minipage}
\hfill
\begin{minipage}[a]{0.5\linewidth}
\centering
\includegraphics[width=8.6cm]{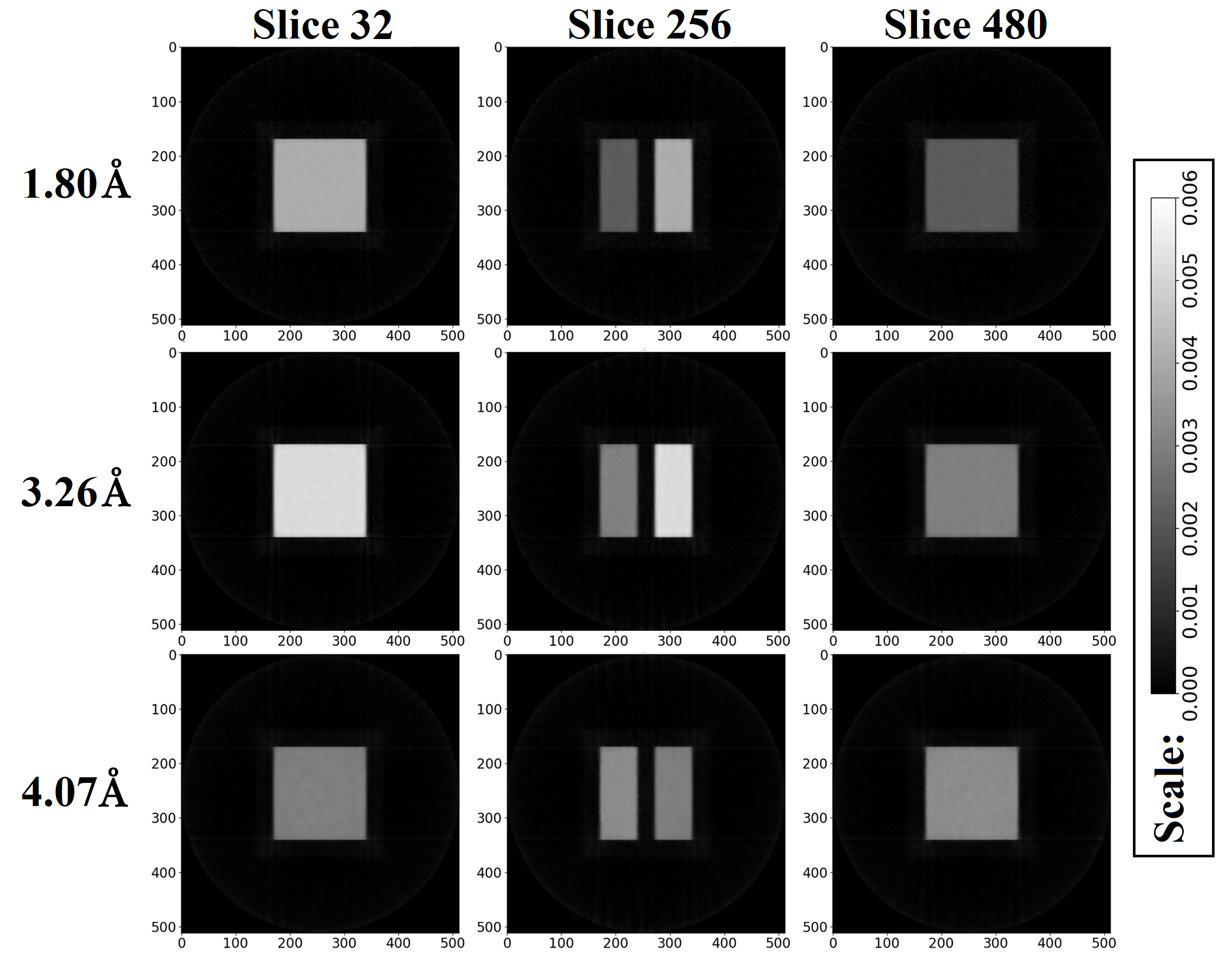}
\centerline{(d) FHR with MBIR}\medskip  
\end{minipage}
\vspace{-0.5cm}
\caption{Hyperspectral reconstruction for simulated data: 
(a) ground truth, (b) baseline method (DHR), (c) FHR with FBP, and (d) FHR with MBIR.
Even with FBP, FHR reconstructions are significantly less noisy compared to DHR, which highlights the effectiveness of subspace extraction in noise reduction.
The use of MBIR further improved FHR reconstructions.}
\label{fig:recon_hyper_sim}
\end{figure*}

The measured dataset was acquired using the Spallation Neutrons and Pressure (SNAP) diffractometer beamline at Oak Ridge National Laboratory (ORNL) \cite{mason2006sns, song2017sns}, which has a wavelength resolution of $\frac{\Delta \lambda}{\lambda}=0.002$.
Similar to the simulation, the dataset covered a wavelength range of 1.5 to 4.5 \AA.
The physical sample was formed from powdered Ni and powdered Cu in an Al structure.
Ni and Cu were in powdered form to avoid internal texture and potential residual strain, ensuring a more controlled experimental setup.
The external dimensions of the physical phantom were $20\times5\times5$ mm, and the data were collected using a $28\times28$ mm TOF microchannel plate Timepix detector \cite{watanabe2017tof}.
However, as the sample was glued to the holder at the top and bottom, we excluded data from the top and bottom 50 rows of the detector.

The associated parameters for both datasets are specified in Table~\ref{table:data_sim_real}.
It is important to note that $N_m$ must be known for FMD implementation.
However, FHR does not require $N_m$, but a rough idea of $N_m$ can help to select a suitable $N_s$.

\begin{table}[ht!]
\begin{center}
\caption{Parameters for Simulated \& Measured Data.}
\label{table:data_sim_real}
\begin{tabular}{|c|c|c|c|c|c|c|c|}
  \hline
  Dataset & $N_r$ & $N_c$ & $N_k$ & $N_v$ & $N_m$ & $N_s$ & $N_q$\\
  \hline
  Simulated & 512 & 512 & 1200 & 32 & 3 & 9 & 12\\
  \hline
  Measured & 412 & 512 & 1200 & 27 & 3 & 9 & 10\\
  \hline
\end{tabular}
\end{center}
\end{table}

Along with our own algorithms, we applied the following baseline methods to the datasets for comparison:
\begin{itemize}
\item \textbf{DHR:} DHR performs $N_k$ individual 3D reconstructions, one for each wavelength bin in $p$.
We implemented DHR using FBP, as it is a more practical choice when dealing with a large number of reconstructions and represents the common practice in this field. 
\item \textbf{RDMD:} RDMD first performs $N_k$ FBP reconstructions, similar to DHR.
From these reconstructions, it then calculates an $N_k$ dimensional vector mean for each material region, representing the material's $\mu$-spectrum.
Once the entire set of $\mu$-spectra $D^m$ is computed, RDMD uses it to estimate the set of material projection views $V^m$ from $p$.
Finally, it performs FBP reconstruction on $V^m$ to obtain the reconstructed materials $x^m$.
Notice that, instead of directly estimating $x^m$ from $x^h$, we return to the sinogram domain to compute $V^m$ and then reconstruct $x^m$.
This approach is more efficient because $x^h$ is much larger than 
$p$, making space-domain processing significantly slower and more memory-intensive.
\end{itemize}

\subsection{Simulated Data Results}
\label{ssec:results_sim}

Figure~\ref{fig:recon_hyper_sim} presents hyperspectral reconstructions for the simulated data.
Figure~\ref{fig:recon_hyper_sim}(a) shows selected ground truth (GT) slices at different wavelength bins, while Figure~\ref{fig:recon_hyper_sim}(b) presents corresponding reconstructions using the baseline method (DHR).
Figure~\ref{fig:recon_hyper_sim}(c) and~(d) display reconstructions for the same slices \& wavelength bins using FHR with FBP and MBIR, respectively.
Notably, FHR with FBP produced reconstructions with significantly less noise compared to DHR, despite not using MBIR.
This highlights the independent contribution of the subspace extraction procedure in performing effective noise reduction.
FHR with MBIR further enhanced the reconstructions by achieving additional noise and artifact suppression.

Table~\ref{table:hr_sim} shows a quantitative performance comparison between DHR and FHR for the simulated data.
We see that FHR methods were dramatically faster than DHR.
The table also presents average SNR values for the reconstructions computed by:
\begin{equation} \label{eq:snr_fhr}
    SNR_{FHR} = 10\log \left( \frac{1}{N_k N_m} \sum_{k=1}^{N_k} \sum_{m=1}^{N_m}\left(\frac{\mu_{m,k}^{signal}}{\sigma_k^{noise}}\right)^2 \right) \ ,
\end{equation}
where $\mu_{m,k}^{signal}$ is the signal mean computed from a $20 \times 20$ region inside the $m^{th}$ material at $k^{th}$ wavelength bin.
$\sigma_k^{noise}$ is the noise standard deviation computed from a $20 \times 20$ background region (outside the sample) at $k^{th}$ wavelength bin.
We can observe substantial improvements in SNR values with both FHR methods.

\begin{figure}[t!]
\centering
\includegraphics[width=0.83\linewidth]{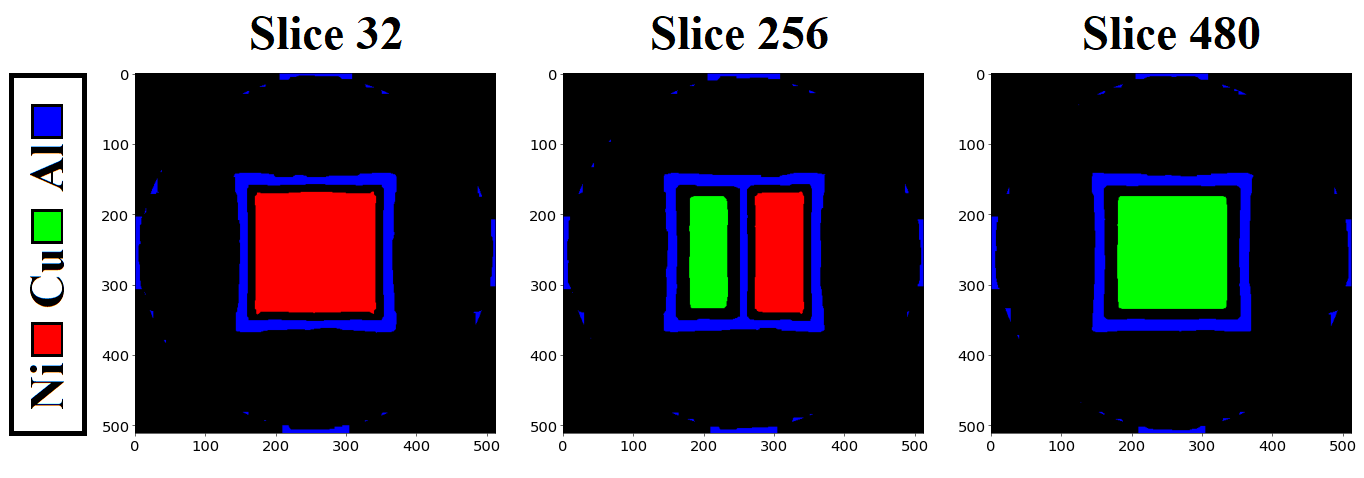}
\caption{Segmented homogeneous material regions used in unsupervised FMD for simulated data.
While the segmentation of Ni and Cu is mostly accurate, Al segmentation shows some mislabeled pixels.}
\label{fig:segment_SIM}
\end{figure}

\begin{figure}[b!]
\centering
\includegraphics[width=0.86\linewidth]{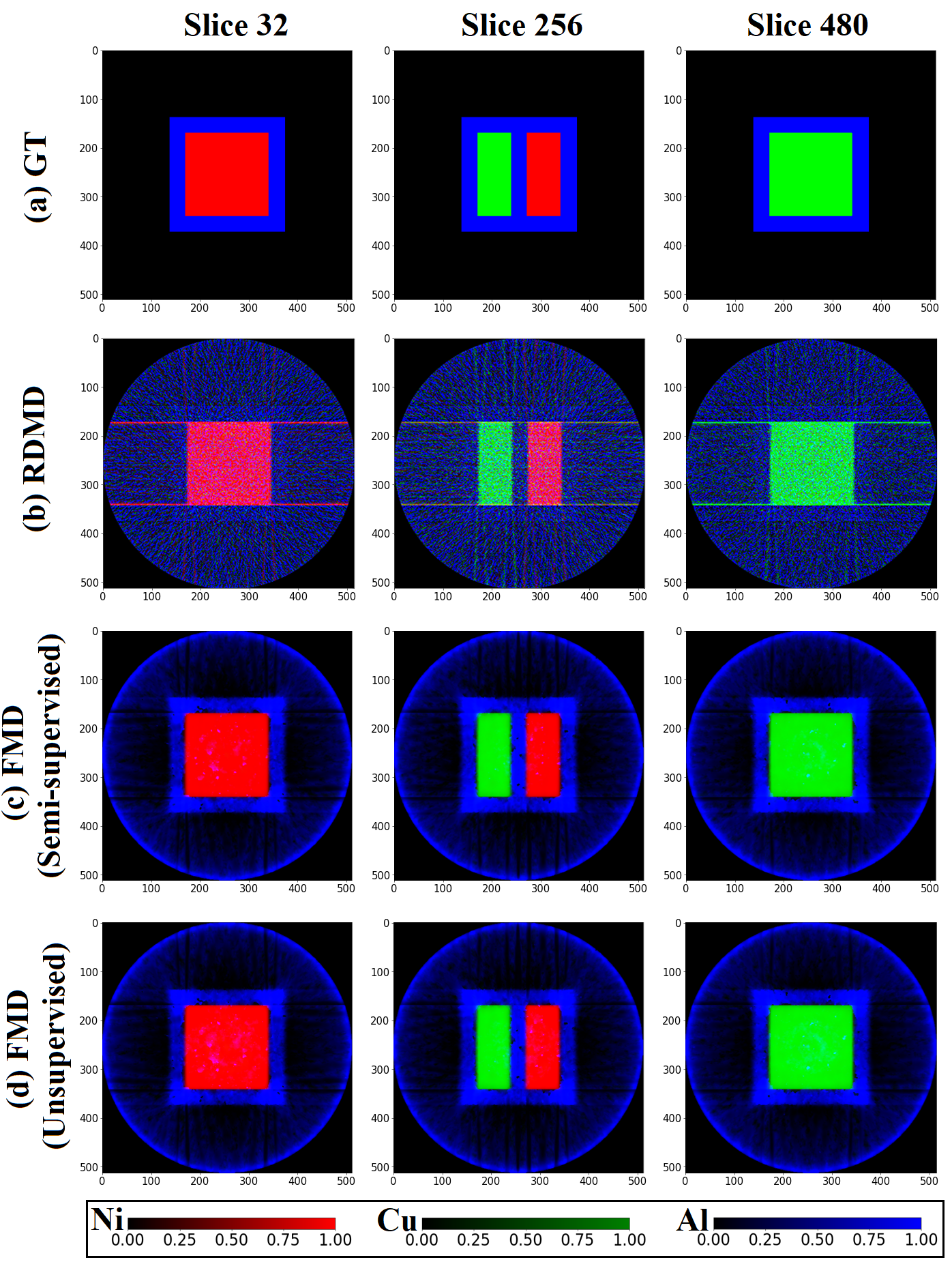}
\caption{Material reconstruction for simulated data: 
Selected slices from reconstructed Ni, Cu, and Al: (a) ground truth, (b) estimated by the baseline method (RDMD), (c) estimated by semi-supervised FMD, and (d) estimated by unsupervised FMD.
The reconstructions represent volume fractions, with values ranging from 0 to 1.
Both semi-supervised and unsupervised FMD reconstructions exhibit significantly lower noise levels than RDMD.}
\label{fig:recon_SIM}
\end{figure}

\begin{table}[ht!]
\begin{center}
\caption{Hyperspectral Reconstruction for Simulated Data.}
\label{table:hr_sim}
\begin{tabular}{|c|c|c|}
  \hline
  Algorithm & Computation Time & SNR\\
  \hline
   Baseline Method (DHR) & 817.44 min & -2.33 dB\\
  \hline
   FHR with FBP & {\bf 23.62 min} & 26.34 dB\\
  \hline
   FHR with MBIR & 62.96 min & {\bf 42.13 dB}\\
  \hline
\end{tabular}
\end{center}
\end{table}

Figure~\ref{fig:segment_SIM} illustrates the homogeneous material regions in the simulated data segmented by FMD for the unsupervised mode.
Notice that the segmentation of Ni and Cu is mostly accurate, while Al segmentation shows some mislabeled pixels in the background.
This phenomenon can be attributed to the weak presence of Al in neutron radiographs, making it difficult to differentiate Al from the background.

\begin{figure}[t!]
\centering
\centerline{\includegraphics[width=4.6cm]{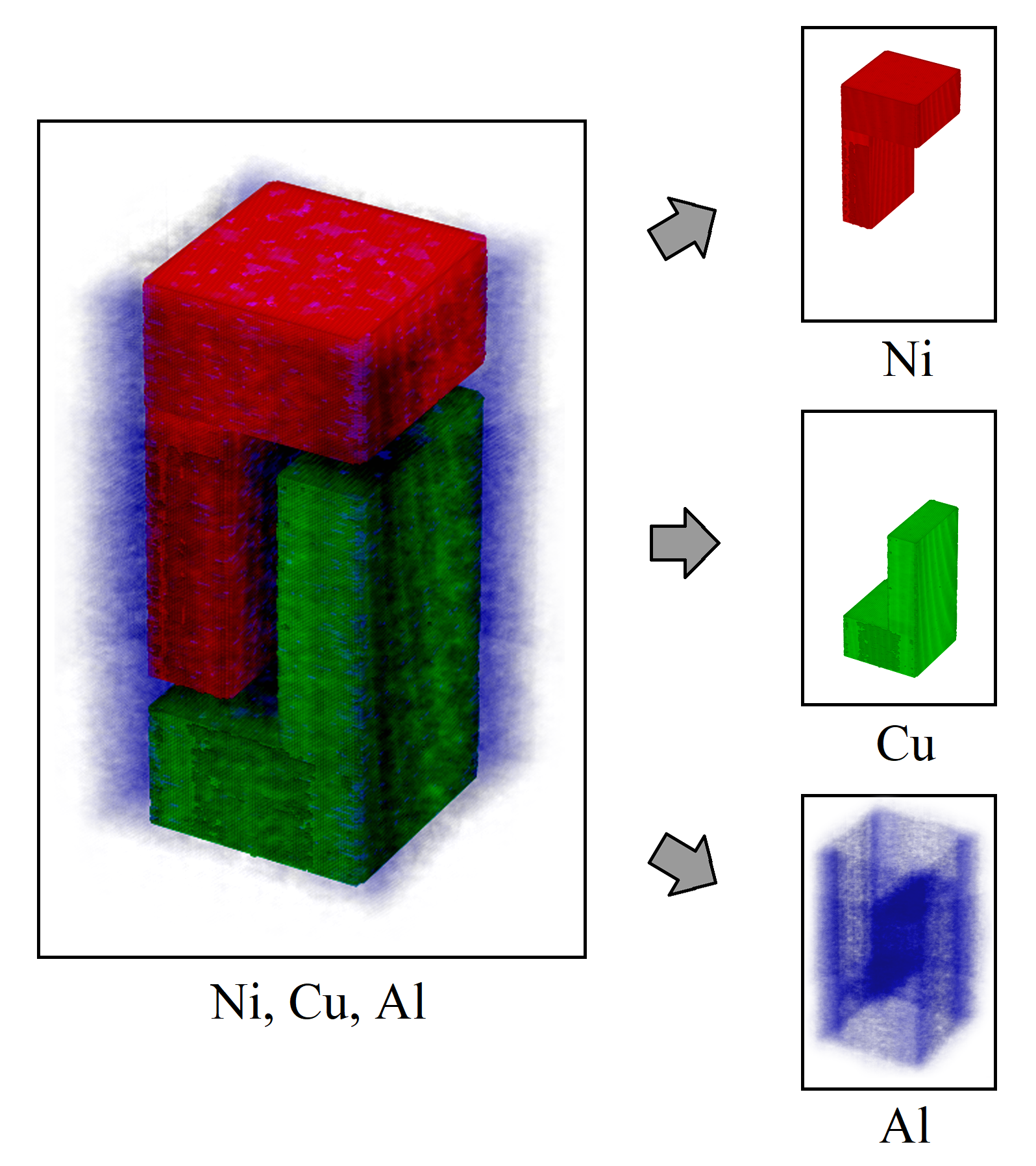}}
\vspace{-0.2cm}
\caption{Material reconstruction for simulated data:  
3D visualization of the reconstructed Ni, Cu, and Al estimated by unsupervised FMD.}
\label{fig:recon_3D_SIM}
\end{figure}

Figure~\ref{fig:recon_SIM} illustrates material reconstructions for the simulated data.
Figure~\ref{fig:recon_SIM}(a) shows several ground truth slices for the material simulation.
Figure~\ref{fig:recon_SIM}(b) shows the baseline RDMD material reconstructions for the same slices.
Figure~\ref{fig:recon_SIM}(c) and~(d) show the associated FMD material reconstructions for the semi-supervised and unsupervised modes, respectively.
Figure~\ref{fig:recon_3D_SIM} shows 3D renderings of the reconstructions for unsupervised FMD.
Comparing FMD reconstructions to the ground truth, we see that the Ni and Cu estimates are accurate for both modes, while Al's estimates show some noise and artifacts.
The circular artifacts in the Al reconstructions arise from the accumulation of residual errors near the boundaries.
While such artifacts are common in CT reconstructions, they are particularly more visible in this case due to the weak presence of Al in neutron radiographs.
However, the RDMD results are much noisier and deviate considerably from the ground truths.

\begin{figure}[t!]
\centering
\includegraphics[width=0.92\linewidth]{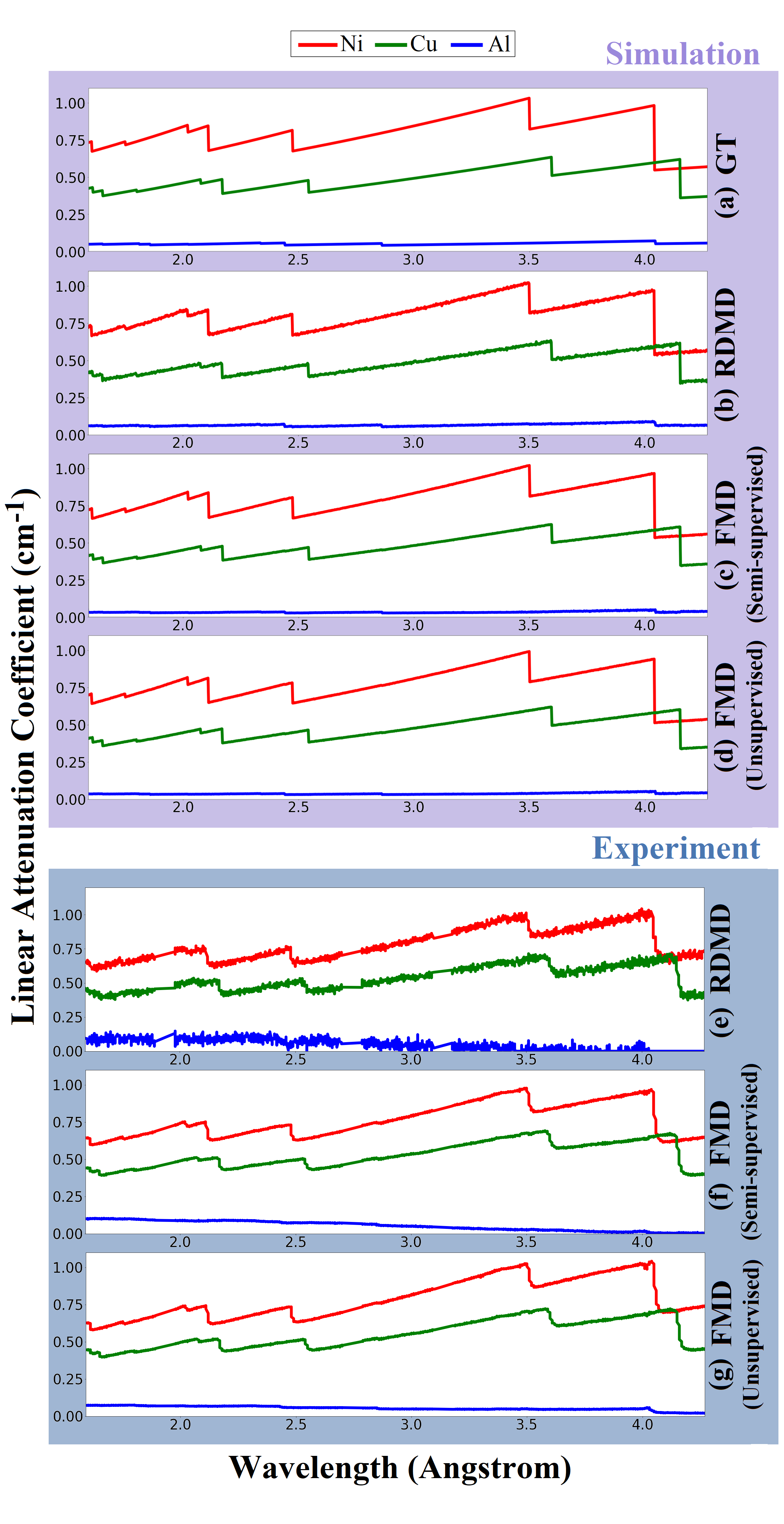}
\vspace{-0.4cm}
\caption{\textbf{$\mu$-spectra for simulated data:} (a) ground truth, (b) using baseline method (RDMD), (c) using semi-supervised FMD, and (d) using unsupervised FMD.
\textbf{$\mu$-spectra for measured data:} (e) using baseline RDMD, (f) using semi-supervised FMD, and (g) using unsupervised FMD.
For both simulated and measured data, FMD methods have estimated $\mu$-spectra with higher SNR than those produced by RDMD.}
\label{fig:LAC_all}
\end{figure}

Figure~\ref{fig:LAC_all}(a) shows the theoretical $\mu$-spectra for Ni, Cu, and Al used in the simulation.
Figure~\ref{fig:LAC_all}(b) shows the spectra estimated using baseline RDMD for the simulated data.
Figure~\ref{fig:LAC_all}(c) and~(d) show the spectra estimated for the simulated data using semi-supervised and unsupervised FMD, respectively.
The FMD estimates of $\mu$-spectra closely match the theoretical ground truths, while RDMD estimates are slightly noisy.

\begin{table}[ht!]
\begin{center}
\caption{Material Decomposition for Simulated Data.}
\label{table:md_sim}
\begin{tabular}{|c|c|c|c|}
  \hline
  \multirow{2}{*}{Algorithm} & Comp. & \multicolumn{2}{c|}{SNR}\\
  \cline{3-4}
  & Time & Material Recon. & $\mu$-Spectra\\
  \hline
  Baseline Method & \multirow{2}{*}{837.62 min} & \multirow{2}{*}{8.10 dB} & \multirow{2}{*}{38.36 dB}\\
  (RDMD) &&&\\
  \hline
  Semi-supervised & \multirow{2}{*}{\bf 79.09 min} & \multirow{2}{*}{\bf 49.28 dB} & \multirow{2}{*}{\bf 50.37 dB}\\
  FMD &&&\\
  \hline
  Unsupervised & \multirow{2}{*}{92.40 min} & \multirow{2}{*}{47.54 dB} & \multirow{2}{*}{48.53 dB}\\
  FMD &&&\\
  \hline
\end{tabular}
\end{center}
\end{table}

Table~\ref{table:md_sim} provides a quantitative performance comparison between RDMD and FMD for the simulated data.
We see that both semi-supervised and unsupervised FMD were faster than RDMD by a significant margin.
Additionally, the table presents average SNR values for the material reconstructions.
The SNR values were computed using the following:
\begin{equation} \label{eq:snr_fmd}
    SNR_{FMD} = 10\log \left( \frac{1}{N_m} \sum_{m=1}^{N_m}\left(\frac{\mu_{m}^{signal}}{\sigma_m^{noise}}\right)^2 \right) \ ,
\end{equation}
where $\mu_{m}^{signal}$ is the signal mean computed from a $20 \times 20$ material region inside the $m^{th}$ material reconstruction.
$\sigma_m^{noise}$ is the noise standard deviation computed from a $20 \times 20$ background region (outside the sample) in the $m^{th}$ material reconstruction.
Using a similar approach, we computed the SNR values for the $\mu$-spectra.
Both semi-supervised and unsupervised FMD achieved significantly higher SNR values compared to RDMD.

\begin{figure}[b!]
\begin{minipage}[a]{0.99\linewidth}
\centering
\includegraphics[width=8.6cm]{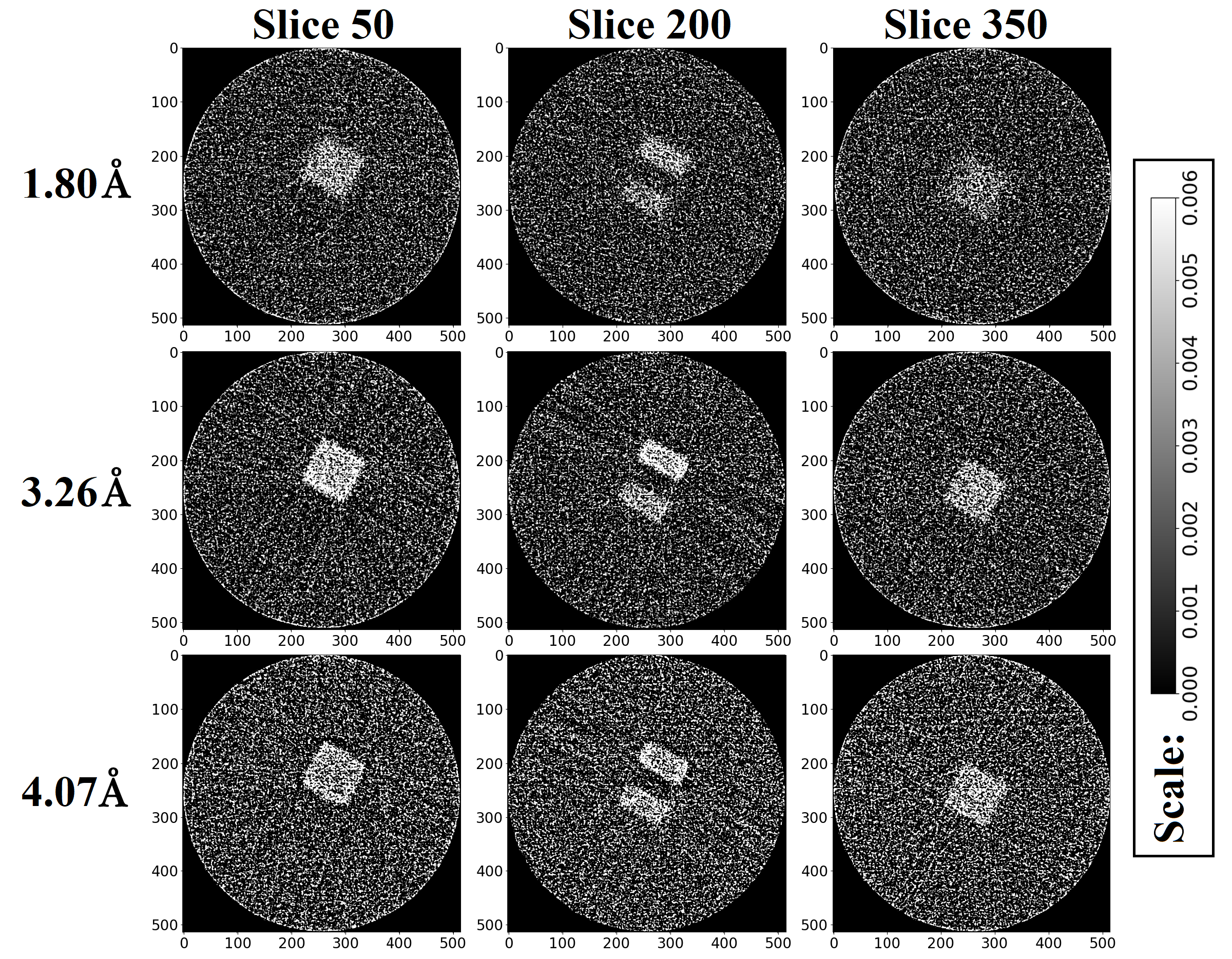}
\centerline{(a) Baseline Method (DHR)}\medskip  
\end{minipage}
\vfill
\begin{minipage}[a]{0.99\linewidth}
\centering
\includegraphics[width=8.6cm]{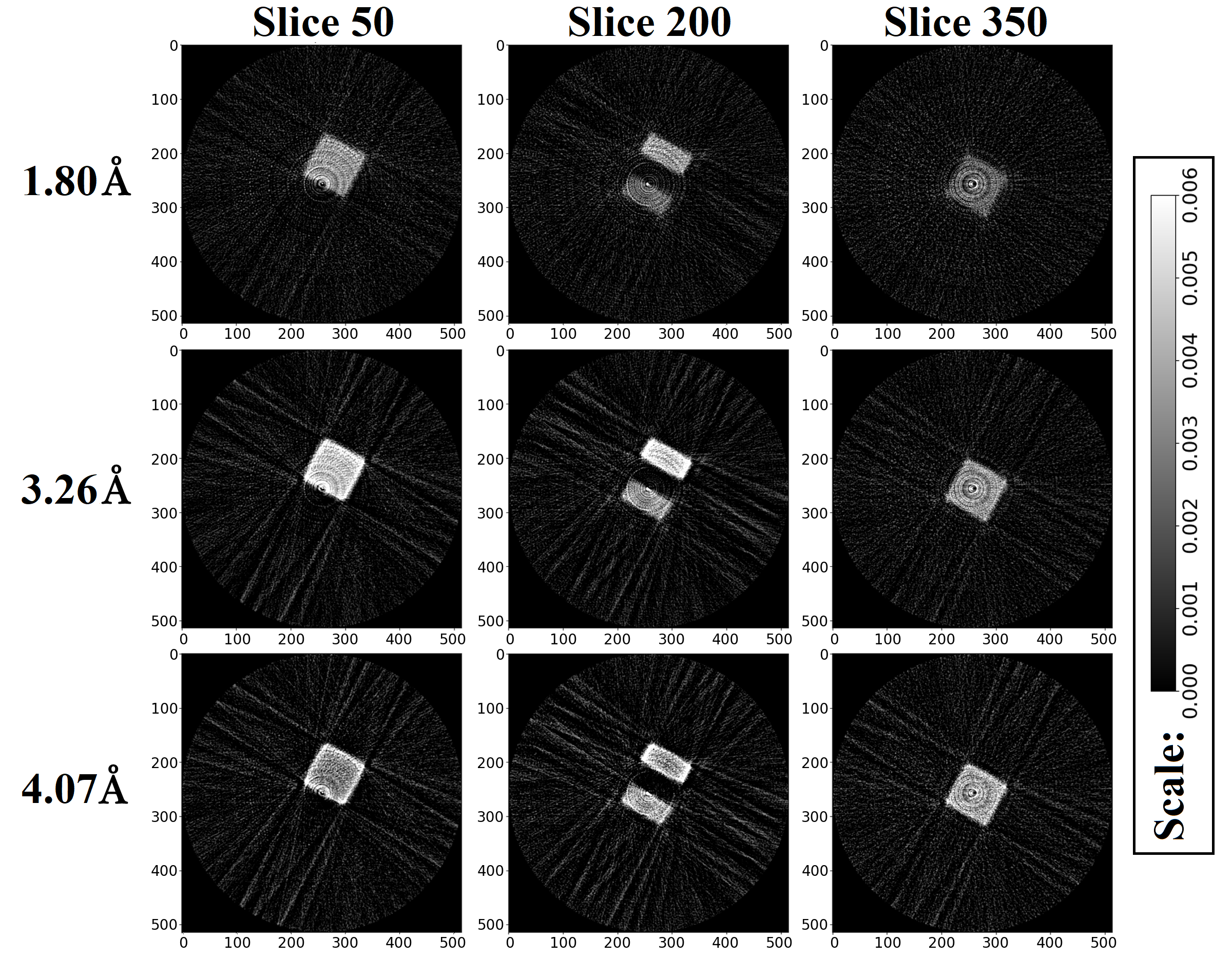}
\centerline{(b) FHR with FBP}\medskip  
\end{minipage}
\vfill
\begin{minipage}[a]{0.99\linewidth}
\centering
\includegraphics[width=8.6cm]{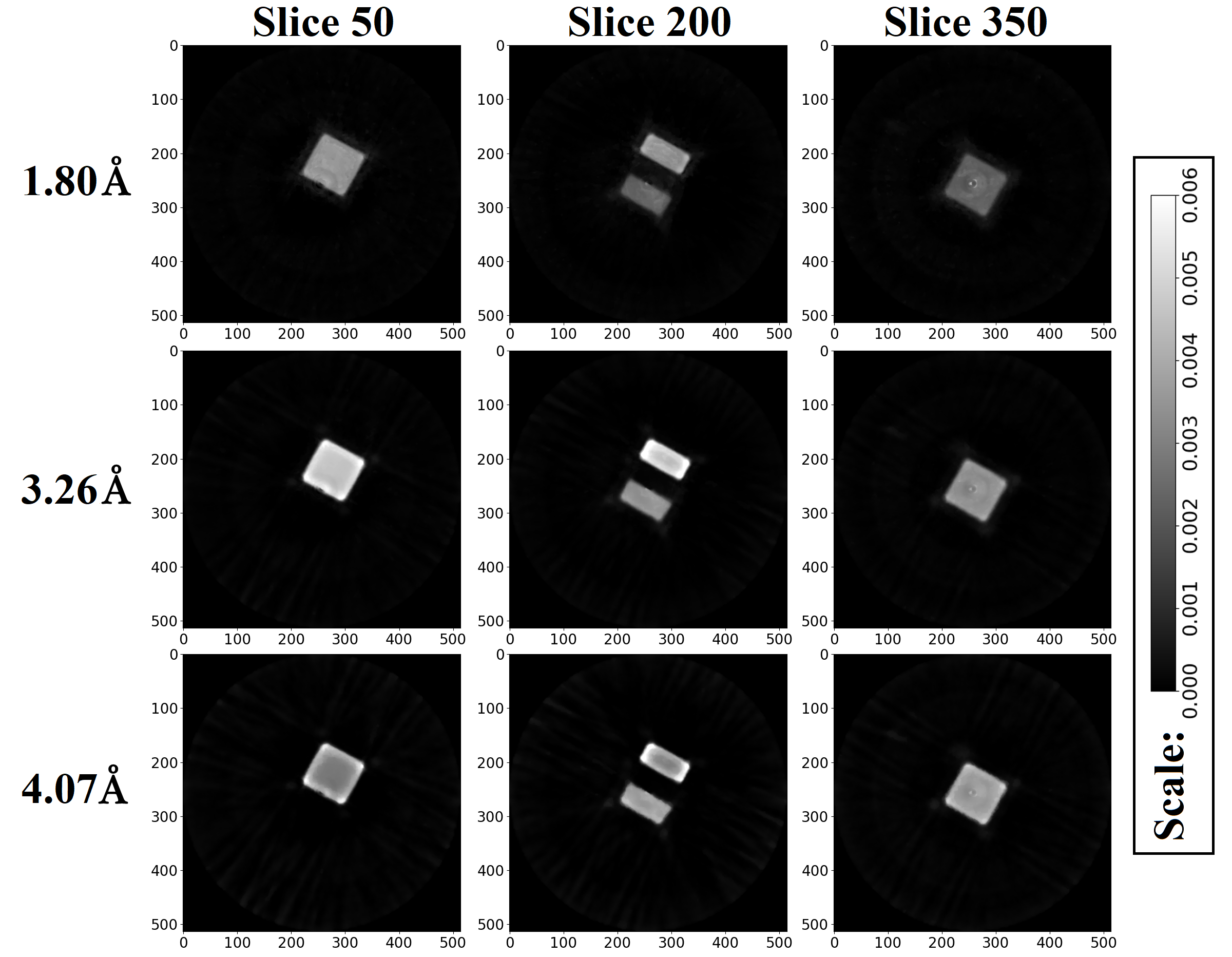}
\centerline{(c) FHR with MBIR}\medskip  
\end{minipage}
\vspace{-0.15in}
\caption{Hyperspectral reconstruction for measured data:
(a) baseline method (DHR), (b) FHR with FBP, and (c) FHR with MBIR.
Both FHR-FBP and FHR-MBIR have produced reconstructions with significantly less noise compared to DHR.
However, FHR-MBIR reconstructions are better than FHR-FBP as expected.}
\label{fig:recon_hyper_real}
\end{figure}

\subsection{Measured Data Results}
\label{ssec:results_real}

Figure~\ref{fig:recon_hyper_real} presents hyperspectral reconstructions for the measured data.
Figure~\ref{fig:recon_hyper_real}(a) shows a selection of reconstructed slices at different wavelength bins using DHR.
Figure~\ref{fig:recon_hyper_real}(b) and~(c) display similar reconstructions using FHR with FBP and MBIR, respectively.
As observed previously, the FHR reconstructions have substantially reduced noise and artifacts relative to DHR.
However, cupping artifacts are present in reconstructions from all methods, particularly visible at higher wavelengths.
Although the exact reason for cupping is unclear, scattering effects could be a possible cause.

Table~\ref{table:hr_real} provides a quantitative performance comparison between DHR and FHR for the measured data.
Similar to the simulated case, FHR methods remarkably outperformed DHR in both speed and noise suppression.

\begin{table}[ht!]
\begin{center}
\caption{Hyperspectral Reconstruction for Measured Data.}
\label{table:hr_real}
\begin{tabular}{|c|c|c|}
  \hline
  Algorithm & Computation Time & SNR\\
  \hline
  Baseline Method (DHR) & 634.06 min & -2.20 dB\\
  \hline
  FHR with FBP & {\bf 18.76 min} & 11.31 dB\\
  \hline
  FHR with MBIR & 60.31 min &{\bf 35.64 dB}\\
  \hline
\end{tabular}
\end{center}
\end{table}

\begin{figure}[b!]
\centering
\includegraphics[width=0.83\linewidth]{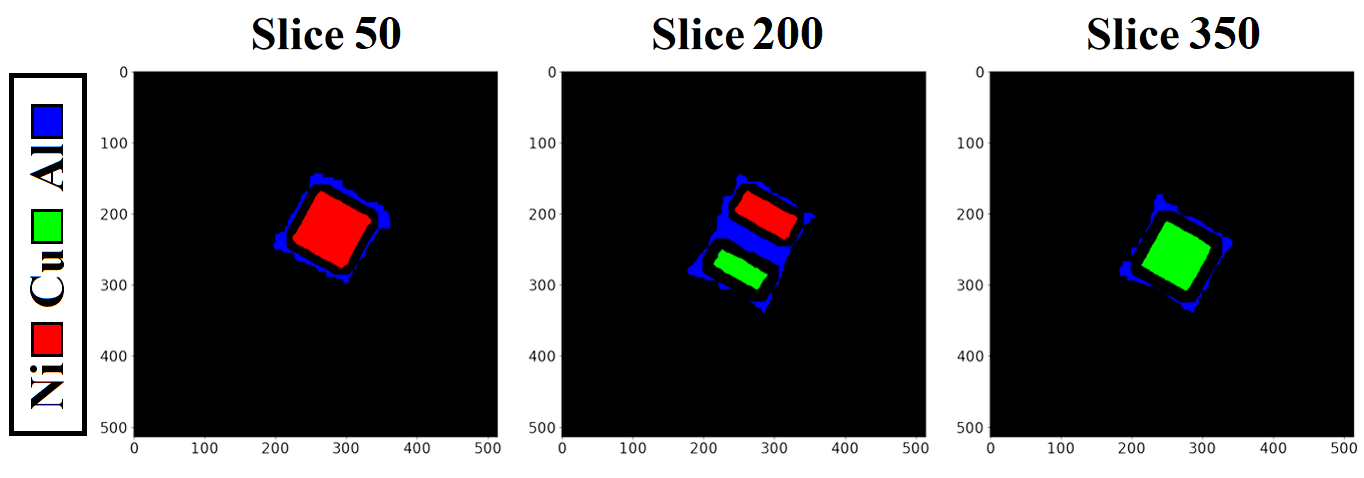}
\vspace{-0.2cm}
\caption{Segmented homogeneous material regions used in unsupervised FMD for measured data.}
\label{fig:segment_REAL}
\end{figure}

\begin{figure}[t!]
\centering
\includegraphics[width=0.86\linewidth]{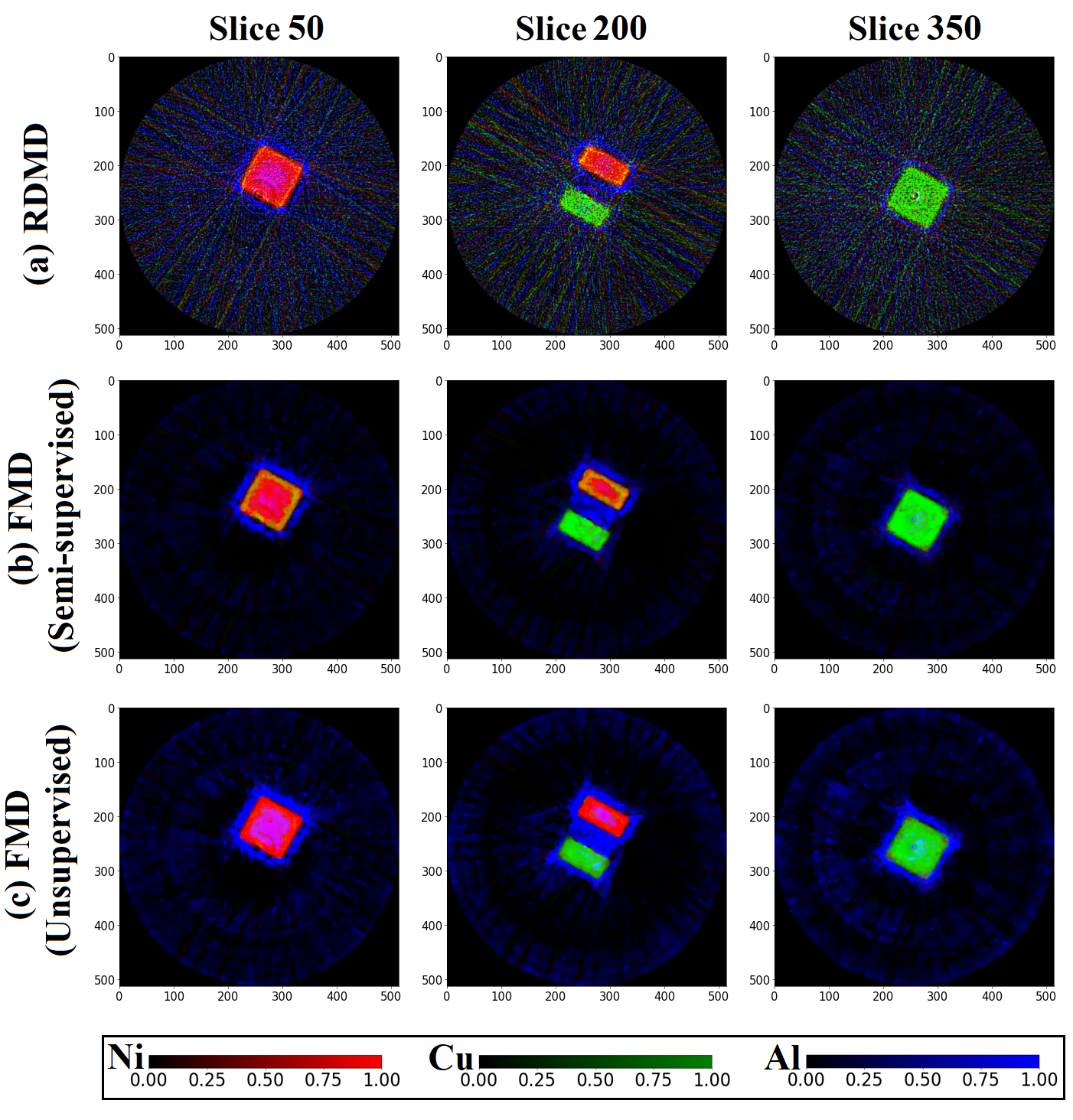} 
\caption{Material reconstruction for measured data: 
Selected slices from reconstructed Ni, Cu, and Al: (a) using the baseline method (RDMD), (b) using semi-supervised FMD, and~(c) using unsupervised FMD.
Both semi-supervised and unsupervised FMD produced better reconstructions with significantly less noise compared to RDMD.}
\label{fig:recon_REAL}
\end{figure}

\begin{figure}[t!]
\centering
\centerline{\includegraphics[width=5cm]{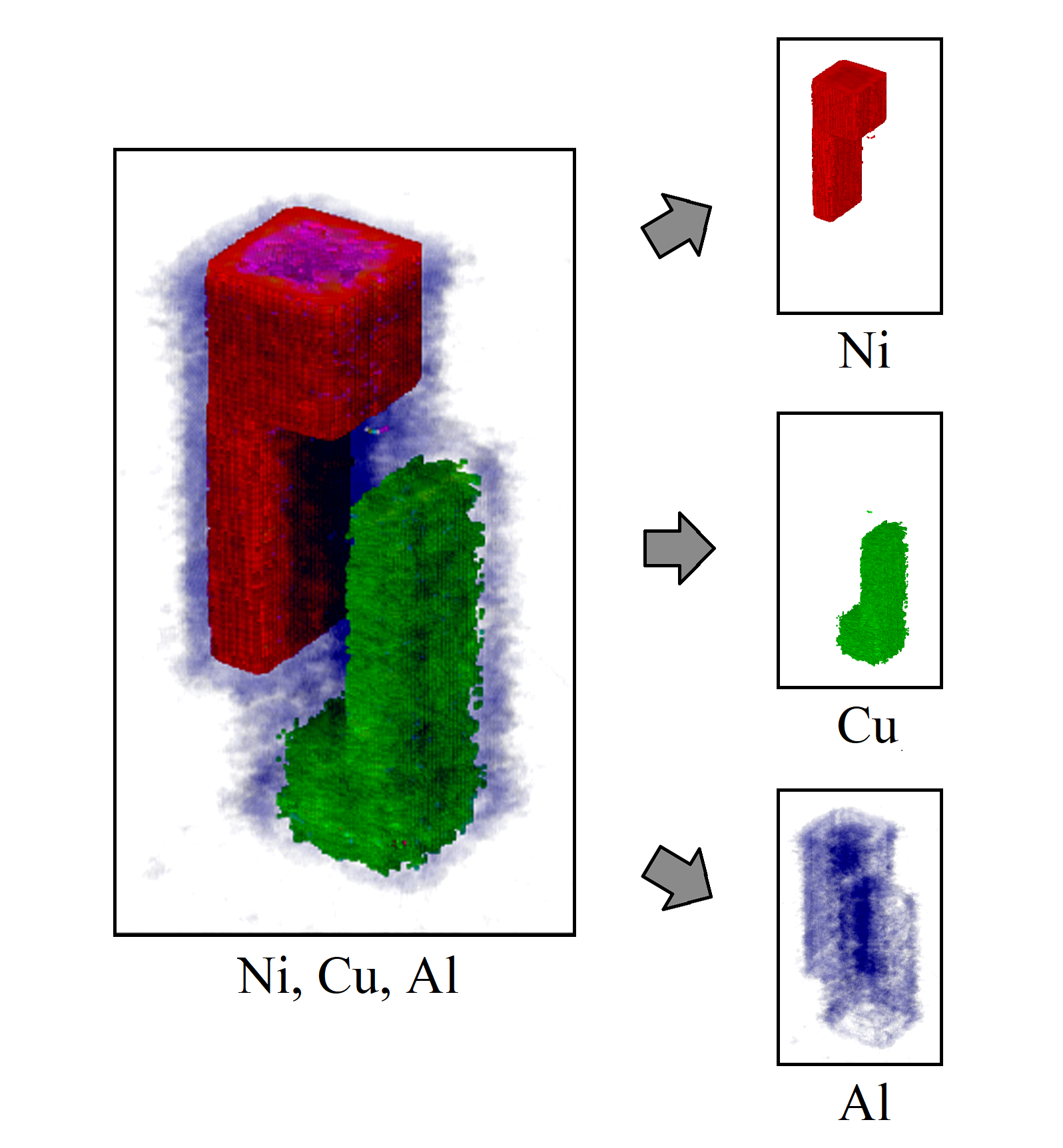}}
\vspace{-0.2cm}
\caption{Material reconstruction for measured data: 
3D visualization of the reconstructed Ni, Cu, and Al estimated by unsupervised FMD.}
\label{fig:recon_3D_REAL}
\end{figure}

Figure~\ref{fig:segment_REAL} illustrates the homogeneous material regions in the measured data segmented by FMD for the unsupervised mode.

Figure~\ref{fig:recon_REAL} illustrates material reconstructions for the measured data.
Figure~\ref{fig:recon_REAL}(a) shows several slices of material reconstructions from the baseline RDMD method, Figure~\ref{fig:recon_REAL}(b) and~(c) show the FMD material reconstructions for the same slices using the semi-supervised and unsupervised modes, and Figure~\ref{fig:recon_3D_REAL} shows the corresponding 3D renderings of the reconstructions for the unsupervised FMD.
Notice that both FMD reconstructions have fewer artifacts with less noise than the RDMD reconstruction.
Also, notice that the unsupervised FMD is somewhat better than the semi-supervised for the Ni and Cu reconstructions, and somewhat worse for the Al.

\begin{table}[t!]
\begin{center}
\caption{Material Decomposition for Measured Data.}
\label{table:md_real}
\begin{tabular}{|c|c|c|c|}
  \hline
  \multirow{2}{*}{Algorithm} & Comp. & \multicolumn{2}{c|}{SNR}\\
  \cline{3-4}
  & Time & Material Recon. & $\mu$-Spectra\\
  \hline
  Baseline Method & \multirow{2}{*}{654.98 min} & \multirow{2}{*}{7.38 dB} & \multirow{2}{*}{20.66 dB}\\
  (RDMD) &&&\\
  \hline
  Semi-supervised & \multirow{2}{*}{\bf 78.50 min} & \multirow{2}{*}{40.44 dB} & \multirow{2}{*}{\bf 44.67 dB}\\
  FMD &&&\\
  \hline
  Unsupervised & \multirow{2}{*}{86.79 min} & \multirow{2}{*}{\bf 41.62 dB} & \multirow{2}{*}{43.54 dB}\\
  FMD &&&\\
  \hline
\end{tabular}
\end{center}
\end{table}

Figure~\ref{fig:LAC_all}(e) shows the $\mu$-spectra estimated for the measured data using baseline RDMD.
Figure~\ref{fig:LAC_all}(f) and (g) show the spectra estimated using semi-supervised and unsupervised FMD, respectively.
Both semi-supervised and unsupervised FMD estimates of $\mu$-spectra have low noise levels.
Additionally, the estimates have Bragg edges that closely match the theoretical calculations from the simulation.
The RDMD estimates of $\mu$-spectra are much noisier compared to FMD.

Table~\ref{table:md_real} provides a quantitative performance comparison between RDMD and FMD for the measured data.
Similar to the simulated case, FMD algorithms were significantly faster than RDMD.
Also, FMD material reconstructions have much higher SNR.

\begin{figure}[b!]
\centering
\centerline{\includegraphics[width=\linewidth]{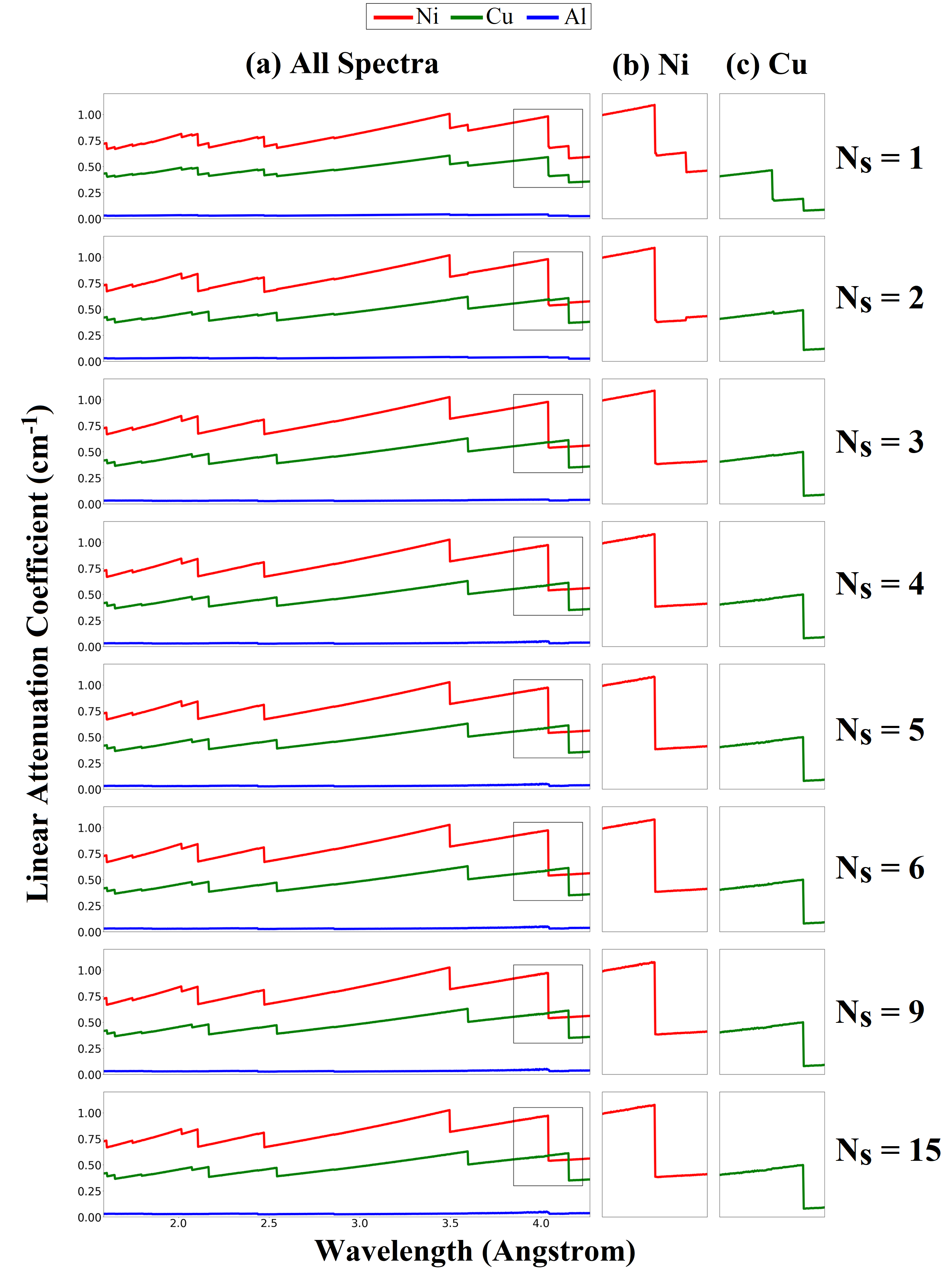}}
\caption{
Simulated data: estimated $\mu$-spectra as a function of the dimension of the intermediate subspace $N_s$.
Each row shows estimated $\mu$-spectra using semi-supervised FMD with the indicated $N_s$ between 1 and 15.
(a) shows the full spectra; (b) and (c) zoom in on the boxed region in (a) for Ni and Cu, respectively. 
The spectra are accurate and nearly identical for $N_s \geq 3$ but show non-physical artifacts for $N_s = 1, 2$.
}
\label{fig:Ns_analysis_sim}
\end{figure}

\subsection{The Dimension of the Intermediate Subspace}
\label{ssec:subspace_dim}

\begin{figure}[t!]
\centering
\centerline{\includegraphics[width=\linewidth]{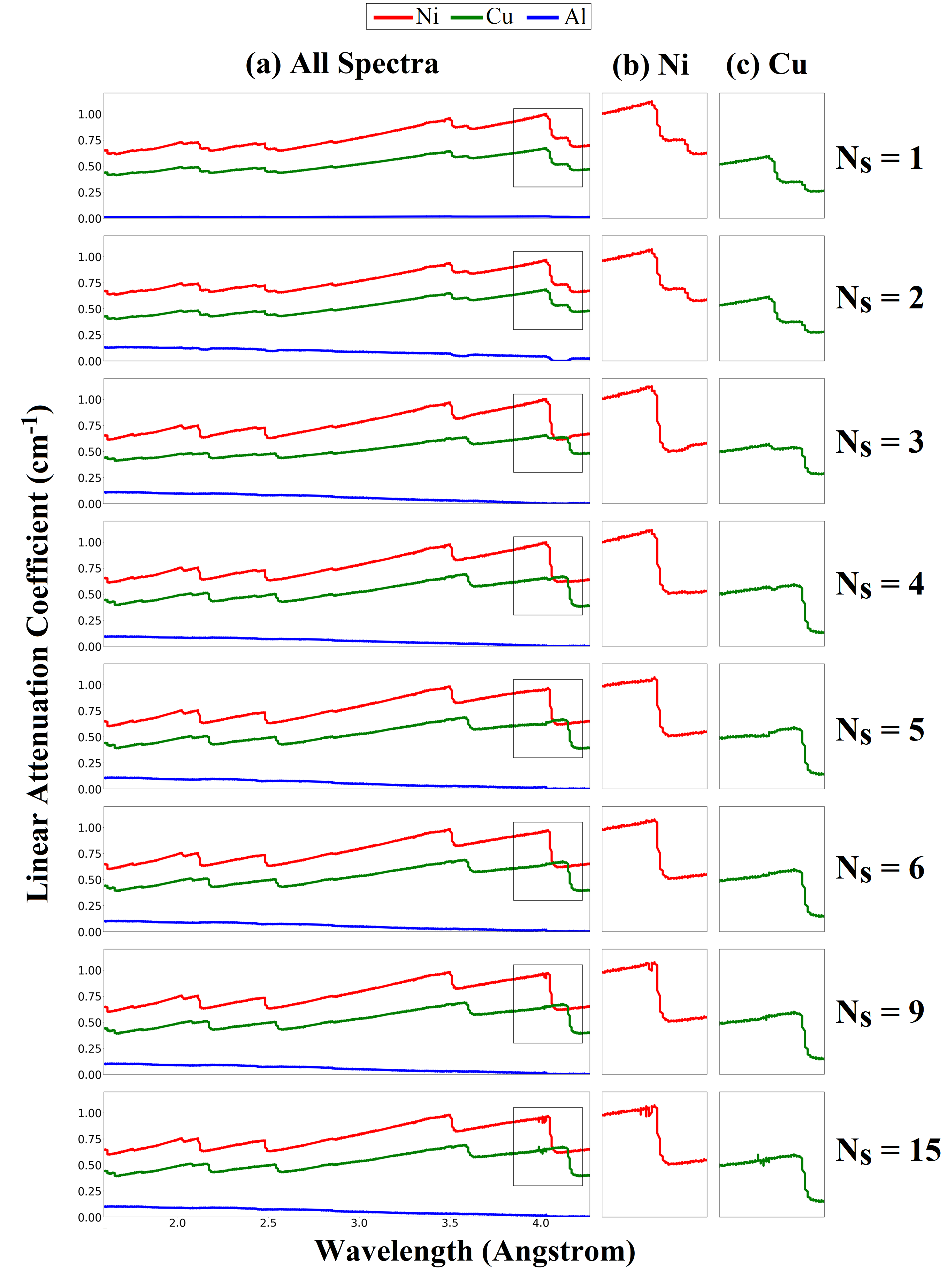}}
\caption{
Measured data: estimated $\mu$-spectra as a function of the dimension of the intermediate subspace $N_s$.
Values of $N_s \leq 3$ show non-physical artifacts similar to those for $N_s = 1, 2$ in Figure 21, with smaller artifacts seen for $N_s=4, 5$.
Values of $N_s$ in the range 6 to 15 give reasonably accurate estimated spectra, with increasingly noisy estimates as $N_s$ increases.
}
\label{fig:Ns_analysis_real}
\end{figure}

In this section, we investigate the best choice for $N_s$, the dimension of the intermediate subspace.
To better understand the effect of $N_s$ on algorithm performance, we conducted an experiment on both datasets using semi-supervised FMD, where we varied the value of $N_s$ from 1 to 15 and observed the estimated $\mu$-spectra.

Figure~\ref{fig:Ns_analysis_sim} shows the estimated $\mu$-spectra for simulated data using different values of $N_s$.
Each row here represents $\mu$-spectra estimations for a single $N_s$.
(a) shows the full $\mu$-spectra, while (b) and (c) provide a closer look at the spectra inside the boxed region in (a) for Ni and Cu, respectively.
Notice that for $N_s \geq 3$ the $\mu$-spectra are accurately reconstructed; whereas for $N_s<3$ the reconstructed $\mu$-spectra have defects.
So, for the simulated dataset, $N_s=N_m=3$ (i.e., $\beta=1$) appears to be the best choice.

Figure~\ref{fig:Ns_analysis_real} shows the estimated $\mu$-spectra for measured data using different values of $N_s$.
Notice that for $N_s \geq 6$ the $\mu$-spectra are accurately reconstructed; whereas for $N_s<6$ the reconstructed $\mu$-spectra have defects.
However, if $N_s$ is too large, then the SNR of the estimated $\mu$-spectra slightly decreases.
So for the measured dataset, $N_s=6$ (i.e., $\beta=2$) appears to be the best choice.

Ideally, the region between any two consecutive Bragg edges in a $\mu$-spectrum is linear.
Thus, an accurately estimated spectrum should have data points between two Bragg edges that closely fit a straight line.
So, for a quantitative evaluation, we computed the ``spectral NRMSE'' metric as the normalized root mean square error (NRMSE) between the estimated $\mu$-spectrum points and a corresponding straight-line fit between the two largest consecutive Bragg edges.
This was done separately for Ni and Cu, and the values were then averaged.

\begin{figure}[t!]
\centering
\centerline{\includegraphics[width=0.8\linewidth]{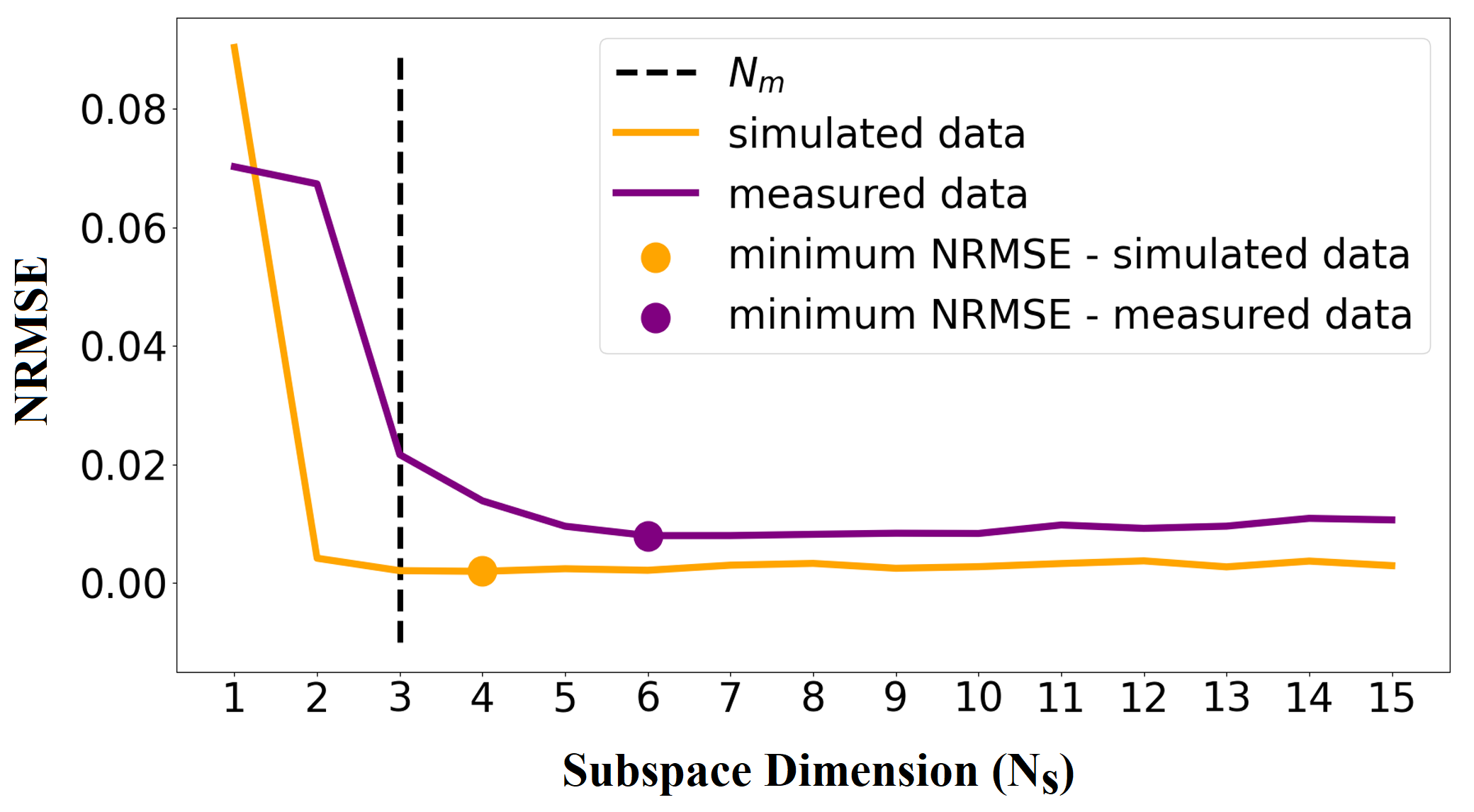}}
\caption{
Spectral NRMSE values as a function of $N_s$.
Using the spectral estimates from Figs~\ref{fig:Ns_analysis_sim} and~\ref{fig:Ns_analysis_real}, we select the region between the two largest consecutive Bragg edges and find the NRMSE between the estimated $\mu$-spectrum and the corresponding straight-line fit.
The NRMSE values for Ni and Cu are averaged and plotted versus $N_s$.
The orange line represents simulated data; the purple line represents measured data.
The dots on each curve denote the $N_s$ values corresponding to the minimum NRMSE.
The vertical dashed line shows the number of materials $N_m$ for both datasets.
Consistent with Figs~\ref{fig:Ns_analysis_sim} and~\ref{fig:Ns_analysis_real}, we see that the fit for simulated data is essentially insensitive to $N_s$ for $N_s \geq 3$, while the fit for measured data is best at $N_s = 6$ but nearly flat over the range 6 to 10.
}
\label{fig:ns_vs_nrmse}
\end{figure}

Figure~\ref{fig:ns_vs_nrmse} shows the spectral NRMSE values as a function of the dimension of the intermediate subspace $N_s$ for both simulated and measured data.
In both cases, the minimum NRMSE was achieved for $N_s > N_m$, and the NRMSE remained relatively stable for larger values of $N_s$.
So, these results indicate that it is better to choose $N_s>N_m$ (i.e., $\beta>1$) when the best choice is unknown.

\section{Conclusion}
\label{sec:conclusion}
We present fast hyperspectral reconstruction (FHR) and fast material decomposition (FMD) algorithms for fast and accurate analysis of polycrystalline material using HSnCT.
The subspace decomposition procedure in both algorithms dramatically reduces data dimensionality and spectral noise, allowing them to operate up to ten times faster and produce more accurate results compared to traditional approaches.
The dramatic reduction in dimensionality also allows for the use of more computationally expensive MBIR reconstruction methods, which further reduce artifacts and improve SNR, particularly in the sparse view case.
While our algorithms have been designed for hyperspectral neutron tomography, similar methods may be useful for other hyperspectral tomography applications.

\section*{Acknowledgment}
\label{sec:acknowledgment}
C. Bouman was partially supported by the Showalter Trust. 
This research used resources at the Spallation Neutron Source, a DOE Office of Science User Facility operated by the Oak Ridge National Laboratory.
The beam time was allocated to the Spallation Neutrons and Pressure Diffractometer (SNAP) instrument on proposal number IPTS-26894.

\appendix

\subsection{Offset Correction}
\label{appendix:off_corr}
During an experiment, a mismatch in neutron dosage rates can happen between the object scan and the blank scan (open-beam), affecting the data normalization process.
In an ideal situation, $y^o$ would have the same neutron dosage rates as $y$.
So, ideally, $p$ can be expressed as
\begin{equation} \label{eq:proj_ideal}
    \nonumber
    p_{v,r,c,k} = -\log\left( \frac {y_{v,r,c,k}} {y_{r,c,k}^o} \right) \ .
\end{equation}
However, due to experimental inaccuracies and instrumental defects, the dosage rates may vary for each view and wavelength bin.
So, we can compensate for the mismatch using a view and wavelength-dependent factor $\alpha \in \mathbb{R}^{N_v \times N_k}$.
Then, $p$ can be expressed as
\begin{align} \label{eq:off_corr_proof}
    p_{v,r,c,k} & = -\log\left( \frac {\alpha_{v,k} y_{v,r,c,k}} {y_{r,c,k}^o} \right) \\
    & = -\log\left( \frac {y_{v,r,c,k}} {y_{r,c,k}^o} \right) - \log(\alpha_{v,k}) \\
    & = -\log\left( \frac {y_{v,r,c,k}} {y_{r,c,k}^o} \right) - b_{v,k} \ , 
\end{align}
where $b = \log(\alpha)$.
The computation of each $b_{v, k}$ is given by
\begin{equation} \label{eq:back_offset}
b_{v,k} = \frac{1}{\vert B \vert} \sum_{(r,c) \in B} {-\log\left( \frac {y_{v,r,c,k}} {y_{r,c,k}^o} \right)} \ .
\end{equation}
$B$ is a set of detector pixel coordinates $(r,c)$ in the background region, where $p_{v,r,c,k}$ is expected to be 0.
$\vert B \vert$ is the number of $(r,c)$ pairs in $B$.

\subsection{$\mu$-Spectra \& Subspace Basis Relationship}
\label{appendix:rel_sub_mat_basis}
From equation \eqref{eq:mat_recon_nmf}, the relationship between the material and subspace reconstructions can be defined as:
\begin{equation} \label{eq:sub_mat_rel}
     x^s = x^m T.
\end{equation}
Taking a linear projection operator $A$ on both sides of the equation \eqref{eq:sub_mat_rel}, we have
\begin{align} \label{eq:mat_sub_bas_proof_1}
    & Ax^s = A(x^m T) \\
    & V^s = (Ax^m) T  \ .
\end{align}
Now, from equation \eqref{eq:porj_matd_rel} and \eqref{eq:sub_extract_solve}, we have
\begin{align} \label{eq:mat_sub_bas_proof_2}
    & (Ax^m) (D^m)^\top = V^s (D^s)^\top \\
    & (Ax^m) (D^m)^\top = (Ax^m) T (D^s)^\top \\
    & (D^m)^\top = T (D^s)^\top \\
    \nonumber	
    & D^m = D^s T^\top  \ . 
\end{align}

\ifCLASSOPTIONcaptionsoff
  \newpage
\fi

\bibliographystyle{IEEEtran}
\bibliography{tci_fhnt}

\end{document}